# Two Egyptian Kings, Shepseskaf and Userkaf, and a solar eclipse (2471 BC)

Amelia Carolina Sparavigna

*Dipartimento di Scienza Applicata e Tecnologia, Politecnico di Torino*

Recently, arXiv published a work by G. Magli about the eclipse of 1 April 2471 BC and a supposed influence of it on the end of the fourth Egyptian dynasty and the beginning of the fifth one. In Magli's arXiv/2412.13640 paper, the eclipse is defined as the 'Shepseskaf eclipse'. Magli considers that this eclipse happened during the reign of the 'last ruler of the 4th dynasty, Shepseskaf'. Some literature is giving the name of further members of this dynasty. Therefore, here we add some references, besides those given in arXiv. We will show that, in this literature, the same eclipse of 2471 BC is linked to king Userkaf of the fifth dynasty. Some observations about chronological data proposed by Magli are also necessary. Since in arxiv/2412.13640 paper we find mentioned by Magli the pyramid of Cheops - that is Akhet Khufu, a name which is generally translated as the Khufu's horizon - and the symbol of the 'solarized' horizon, N27 of the Gardiner's list, it is necessary to stress that the word Akhet was written with the crested ibis, G25, as clearly shown by the Wadi el-Jarf papyri, which are contemporary to Khufu's reign. Mark Lehner, 2001, recognizing a translation problem, rendered Akhet Khufu as the 'Spirit of Khufu'. The use of the crested ibis, in a locution translated as 'dwellers of the horizon' is attested in a Pepi II letter to his vizier Herkhuf, then three centuries after Khufu.

For supplementary material and a detailed discussion of Akhet, see please my "The Egyptian Hieroglyph of the Crested Ibis, from the Cheops' pyramid (Akhet Khufu) to the Akhenaten's Glory of Aten", https://doi.org/10.5281/zenodo.14587114. There I proposed Akhet Khufu as the "Enduring Forever Glory of Khufu".

**Gautschy and coworkers' Egyptian chronology**

Before starting a survey of literature, let us remember the chronology of Egyptian pharaohs as proposed by Gautschy et al., 2017. This is strictly necessary because in the arXiv's test, written by G. Magli, it is told the following: "The bulk of the chronology of ancient Egypt is based on registered astronomical observations of Lunar dates and Heliacal rising of Sirius. These observations … actually furnish astronomical anchors on which specialists can rely in dating the regnal periods of each Pharaoh. Unfortunately, besides a controversial proposal (Gautschy et al. 2017) the first available registrations come from the Middle Kingdom, so no Sothic anchor is as yet available for the Old Kingdom". For what is regarding the Sothic cycle and the Egyptian calendar, see please Sparavigna, 2024, https://arxiv.org/abs/2411.08061, where the reader can find further literature, besides that given by Magli.

Regarding Gautschy and coworkers' article, Magli does not explain what their proposal is, and why is it "controversial" (he does not provide any reference about this controversy). The burden of proof is with Magli, because he mentioned the existence of a problem. However, the reader's convenience allows me to tell that Gautschy and coworkers, 2017, are discussing a "recently discovered inscription on an ancient Egyptian ointment jar [which] mentions the heliacal rising of Sirius. In the time of the early Pharaohs, this specific astronomical event marked the beginning of the Egyptian New Year and originally the annual return of the Nile flood, making it of great ritual importance". The researchers discuss the lunar calendar and provide a detailed list of inscriptions related to the heliacal rising of Syrius. In Puchkov, 2024, the author tells that the "Sothic chronology of the Old Kingdom has not yet been securely established because of the small number of suitable dates from this period". Puchkov considers the same jar discussed by Gautschy and coworkers, adding that "Since key features of the artifact cause dating controversy, it is not possible to benefit from this Sothic date, at least until a photo of the object becomes available". Here, the "controversy" (to my best knowledge, Puchkov is the only author referring to it).

Puchkov: "Although the authors claim that the object originates from the 5th Dynasty [Gautschy et al. 2017, 71], its stylistic features point either to the 6th Dynasty according to Günther and Wellauer's typology, if object Nr. 82 is the best fit, or to the MK [Middle Kingdom] according to Aston's typology, and the use of the term prt Spdt makes it doubtful that the inscription was carved before the MK. The sum of the facts leaves only two possibilities: a) reuse of the 6th Dynasty jar with the addition of an inscription during the MK (unlikely); b) the jar from the MK, but the text was read/interpreted incorrectly (likely). Since key features of the artifact cause dating controversy, it is not possible to benefit from this Sothic date, at least until a photo of the object becomes available".

Let us stress that to explain in a detailed manner why he, Magli, considers the proposal by Gautschy and coworkers a "controversial proposal" is with Magli.

**Chronology**

Let us propose here two chronological lists from Gautschy and coworkers' paper. The following Table is giving the content of Table 3 in Gautschy et al., 2017, proposing the "absolute dates and number of counted regnal years" in Gautschy and coworkers' Low Chronology (left) and High Chronology (right) of the Old Kingdom. Note Thamphitis after Shepseskaf. Shepseskaf was not the last king of the fourth dynasty, according to Gautschy et al.

|  | Low Chronology |  |  | High Chronology |  |
|---|---|---|---|---|---|
| Year | Y1 Pharaoh | Regnal Years | Year | Y1 Pharaoh | Regnal Years |
| 2503 BCE | Khufu | 30 | 2636 BCE | Khufu | 30 |
| 2473 BCE | Djedefre | 9 | 2606 BCE | Djedefre | 9 |
| 2464 BCE | Chefrem | 24 | 2597 BCE | Chefrem | 24 |
| 2440 BCE | Baka | 3 | 2573 BCE | Baka | 3 |
| 2437 BCE | Menkaure | 19 | 2570 BCE | Menkaure | 19 |
| 2418 BCE | Shepseskaf | 6 | 2551 BCE | Shepseskaf | 5 |
| 2412 BCE | Thamphitis | 4 | 2546 BCE | Thamphitis | 2 |
| 2407 BCE | Userkaf | 10 | 2544 BCE | Userkaf | 10 |
| 2398 BCE | Sahure | 19 | 2543 BCE | Sahure | 19 |
| 2379 BCE | Neferirkare | 14 | 2515 BCE | Neferirkare | 14 |

The following screenshot is proposing the Table 4 Gautschy and coworkers (reason to propose a screenshot exists). The caption tells that the Table 4 is giving the "Absolute accession dates of selected pharaohs in different chronological models. From left to right: [Gautschy and coworkers'] Low Chronology; dates given in Hornung, et al., Chronology; Shaw, Oxford History; von Beckerath,

"Chronologie"; the modelled 95% probability C14 accession dates of Dee, et al., "Radiocarbon-based Chronology"; and [Gautschy and coworkers'] High Chronology

| Y1 Pharaoh | Low chronology | Hornung et al. | Shaw | von Beckerath | Dee (C14: 95%) | High chronology |
|---|---|---|---|---|---|---|
| 1 Khufu | 2503 BCE | 2509 BCE | 2589 BCE | 2604/2554 BCE | 2629–2558 BCE | 2636 BCE |
| 1 Shepseskaf | 2418 BCE | 2441 BCE | 2503 BCE | 2511/2461 BCE | 2556–2476 BCE | 2551 BCE |
| 1 Djedkare | 2331 BCE | 2365 BCE | 2414 BCE | 2405/2355 BCE | 2486–2400 BCE | 2468 BCE |
| 1 Unas | 2294 BCE | 2321 BCE | 2375 BCE | 2367/2317 BCE | 2450–2364 BCE | 2432 BCE |
| 1 Pepy I | 2256 BCE | 2276 BCE | 2321 BCE | 2335/2285 BCE | 2399–2310 BCE | 2392 BCE |
| 1 Pepy II | 2201 BCE | 2216 BCE | 2278 BCE | 2279/2229 BCE | --- | 2334 BCE |

In the Appendix 1 of Magli's arXiv, 2024, we find that "in table 1, four among the most accepted chronologies for the Old Kingdom are reported: (1) Baines and Malek (1981), (2) von Beckerath (1997), (3) Shaw (2000), (4) Hornung, Krauss and Warburton (2006) (the asterisks signal a short reign attributed to "Bicheris" in (2) and (3))." Then, Magli is giving the Table according to three references (von Beckerath (1997), (3) Shaw (2000), (4) Hornung, Krauss and Warburton (2006)), which are the same used for their Table 4 by Gautschy and coworkers.

Therefore, let us compare the two Tables, in the case of king Khufu. Table 1 from Magli, arXiv, 2024:

(2) von Beckerath (1997), (3) Shaw (2000), (4) Hornung, Krauss and Warburton (2006)

| King | 1 | 2 | 3 | 4 |
|---|---|---|---|---|
| Khufu | 2551-2528 | 2554–2531 | 2509–2483 | 2589–2566 |

Table 4 from Gautschy et al., "Absolute accession dates of selected pharaohs in different chronological models. …". I referred to this Table in my https://arxiv.org/pdf/2411.08061

| Y1 Pharaoh | Low chronology | Hornung et al. | Shaw | von Beckerath | Dee (C14: 95%) | High chronology |
|---|---|---|---|---|---|---|
| 1 Khufu | 2503 BCE | 2509 BCE | 2589 BCE | 2604/2554 BCE | 2629–2558 BCE | 2636 BCE |

Three references are the same, but the years are different.
3

*For reader's convenience, further discussion about chronology is given in this work at pages 11-12).*

Magli continues: "The length of each single regnal period is subject to some uncertainty on its own, and each regnal period has to be considered within a safety band which in itself is difficult to estimate but cannot be less than, say, ±25 years. It is thus seen that 2471 b.C. can be accommodated practically in all Shepseskaf chronologies. Further, if the *astronomical dating of the Khufu pyramid* that points to **2550** with an uncertainty not greater than **±10 years** for the **first year of Khufu** is accepted, than the "low" and the "high" chronologies of columns (3) and (4) are excluded." Magli does not consider the High and Low chronology given by Gautschy et al. (the jar is "controversial" according to Puchkov).

No reference is given for the astronomical dating of Khufu Pyramid. Gautschy et al., 2017, or Belmonte, 2001? What does it mean that the regnal period has a "safety band", what is the "safety band" in the case of Egyptian chronology? Why, this band cannot be "less than, say, ±25 years"?

**Gautschy and coworkers: Low Chronology Versus High Chronology**

Let us consider a part of Gautschy and coworker's text, which discusses the features of low and high chronologies. In this part of the discussion, please consider that "our" refers to Gautschy and coworkers. "Our Low and High Chronologies are separated by ca. 130 years. The small amount of available astronomical data as well as the documented years of kings do not allow for a justifiable decision to favour one of the two chronological models. One has to state that our Low Chronology with the end of an assumed 54-year reign of Pepy II in 2150 BCE aligns well with the assumption about a short duration of the First Intermediate Period (30 to 60 years) if 143 years are counted for Dynasty 11 and 110–125 years for Dynasty 12 until the Sothic date in year 7 of Senwosret III in 1866 BCE or 1841 BCE. *However, recent radiocarbon data seem to contradict such a low chronology*. On the other hand, the date of Khufu in our model (2504–2475 BCE) coincides with the dating of this Pharaoh suggested by Kate Spence based on the alignment of his pyramid with the help of observations of simultaneous transits of two circumpolar stars (ζ Ursae Minoris and β Ursae Minoris in ca. 2478 BCE). In the *Low Chronology model the jar* with a Sothic date day 1 of the 4th month of the Akhet season (2419–2406 BCE) in Zurich would fall in one of the reigns of *Shepseskaf, Thamphitis or Userkaf* and hence in the transition between Dynasties 4 and 5. *On stylistic grounds an attribution to the later 5th Dynasty would seem better*" (Gautschy and coworkers, mentioning Habicht et al., 2015).

**Queen Chentkaus I**

In the book by Habicht and Habicht of 2022, we can find the following chronology about the reign of Shepseskaf.

Shepseskaf – Regierungjahre [years of reign]: 4 Jahre (Turiner Königspapyrus, nur bruchstückhaft), 7 Jahre (Manetho), 5 Jahre (Gautschy et al. 2017), 5 Jahre (von Beckerath 1997, 159). - Chronologie: 2551-2546 v. Chr. (Gautschy et al., 2017), Um 2510-2500 v. Chr. (Schneider 1996, 492), 2503-2498 v. Chr. (I. Shaw 2003, 482), 2511-2506 v. Chr. (von Beckerath 1997, 159) – Ehefrauen [Wives]: Chent-kau-s I, Bunefer – Nachfolger [Successor]: Thamphitis (Chentkaus I. ?).

The successor is given as Thamphitis, following screenshot, where the name is associated to that of Shepseskaf's wife, Queen Chentkaus I.

**Thamphitis (Königin Chentkaus I.) (?)**

Kurzprofil:
Regierungsjahre: 2 Jahre (von Beckerath 1997, 159)
Chronologie: 2506-2504 v. Chr. (von Beckerath 1997, 159)

See please the discussion about this queen, given by Habicht and Habicht, 2022.

**The last members of the fourth dynasty**

In Gautschy et al., 2017, we find Thamphitis after Shepseskaf. Shepseskaf was not the last king of the fourth dynasty, according to Gautschy et al. Let us stress that Puchkov's "controversy" is regarding a jar (Puchkov, 2024), not the fact that a further member of the fourth dynasty has been mentioned.

https://pharaoh.se is a "website mostly about the kings of ancient Egypt. Information provided "as is". © 2011–2024 Peter Lundström | v. 4.5. Errors & omissions are possible. Always check the sources! Creative Commons Attribution 4.0 License.

Let us start from **Userkaf,** the first king of the fifth dynasty and go back to arrive to **Shepsekaf.**

In https://pharaoh.se/ancient-egypt/pharaoh/userkaf/, we find told that "As the first ruler of the fifth dynasty, Userkaf imitated or honored Djoser by using the same name for his Horus and Nebty names instead of two different ones. He likely married into the extended royal family of the fourth dynasty to gain the throne. His hellenized name by Manetho is Ouserkheres, and his name is present on the New Kingdom king lists of Abydos, Saqqara and Turin. According to a story in the Westcar papyrus, he was the first of three brothers that ruled in succession, but that story has since been disproved." Precedessor Hordjedef, Successor Sahura.

The following screenshot is the chronological table proposed by the site

| Reign of Userkaf | |
| --- | --- |
| AE Chronology | 2435–2429 |
| v. Beckerath | 2479–2471 |
| Shaw | 2494–2487 |
| Dodson | 2392–2385 |
| Allen | 2465–2458 |
| Malek | 2454–2447 |
| Redford | 2513–2506 |
| Turin King List | 7 years |
| Manetho (Africanus) | 28 years |

In https://pharaoh.se/ancient-egypt/pharaoh/hordjedef/ we find the *last member of the fourth dynasty*, *Hordjedef*. "The ninth pharaoh of the Fourth Dynasty, a.k.a. Djedefhor. Hordjedef was a son of Khufu and half-brother of pharaohs Radjedef and Khafra. His name is mentioned on an inscription in Wadi Hammamat, written after the names of Khufu, Radjedef and Khafra, preceding the name of another of his brothers, Baufra. There is no evidence that either Hordjedef or Baufra ruled as a pharaoh, even though only pharaohs' names were written in cartouches during the 4th dynasty. The name appears with an added Ra sign, which implies that the inscription was not contemporary, but inscribed later, likely during the Middle Kingdom." Predecessor is Thamphthis, successor is Userkaf, of the fifth dynasty.

See http://giza.fas.harvard.edu/ancientpeople/669/full/

https://pharaoh.se/ancient-egypt/pharaoh/thamphthis/ Thamphthis is the eighth pharaoh of the Fourth Dynasty, a.k.a. Djedefptah, Ptahdjedef. "Thamphthis is possibly a hellenized version of Djedefptah or Ptahdjedef. Most likely he never ruled the kingdom, but Middle Kingdom tradition seem to venerate his name and somewhere along the line, his name was included among the already then ancient kings. There are no attestations and no archaeological records whatsoever of this king, if he ever existed. The name is only mentioned as the last king in Africanus list of Manetho's fourth dynasty." Predecessor Baufra, Successor Hordjedef.

https://pharaoh.se/ancient-egypt/pharaoh/baufra/ Baufra is the seventh pharaoh of the Fourth Dynasty, a.k.a. Baefra, Baufre, Baka, Bicheris, Bikka. "According to Africanus, Manetho named Bicheres as the successor to Suphis (Khafra) but there is no archaeological evidence for a king by that name. Herodotus and Diodorus writes that after Khufu died, his brothersic Chephren (Khafra) took the throne, but Diodorus correctly adds that it was Khufu's son Chabryen (Radjedef) who ascended the throne. Khafra and Radjedef were in fact brothers, which shows that Herodotus and Diodorus simply misunderstood the relation told by the Egyptian priests, or that there were inconsistencies in the traditions about the fourth dynasty even in ancient Egypt. Maybe the earlier records (on papyrus) were corrupted and only partially readable. …". Predecessor Shepseskaf, Successor Thamphthis.

https://pharaoh.se/ancient-egypt/pharaoh/shepseskaf/ *Shepseskaf* was the sixth pharaoh of the Fourth Dynasty. "According to the archaeological record. Shepseskaf was the last pharaoh of the fourth dynasty. The Turin papyrus and the Saqqara king list is damaged where the last rulers of the dynasty would be placed, and to make matters worse, they detail a different number of pharaohs between Khafra and Userkaf, the first ruler of the fifth dynasty. According to Africanus, Sebercheres was the seventh of eight kings in Manetho's fourth dynasty."

Reing of Shepseskaf, according to the web site:

| Beckerath | Shaw | Dodson | Allen | Arnold | Malek | Redford |
|---|---|---|---|---|---|---|
| 2486-2479 | 2503-2498 | 2396-2392 | 2472-2467 | 2454-2450 | 2460-2456 | 2523-2519 |

Bicheris has his pyramid (Cargnino, 2021). In Budge, 1902, 2013, we find the following:

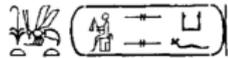

…

Cargnino is also discussing the Shepseskaf tomb (Cargnino, 2021).

The burden to prove that Shepseskaf was the last king of the fourth dynasty is with Magli. In Budge,1902, new edition 2013, we find also what Herodotus reports about the Shepseskaf tomb. "The pyramid here mentioned was undoubtedly built of mud bricks, but that it is to be identified with the Pyramid of Shepses-ka-f, which was called "Qebh" is very unlikely." (Budge, 1902).

Further members of the fourth dynasty are mentioned in https://pharaoh.se . In the web pages of each member of the dynasty, we can find references.

Once more: Onus probandi incumbit ei qui dicit, non ei qui negat.

**Mastaba, not pyramid**

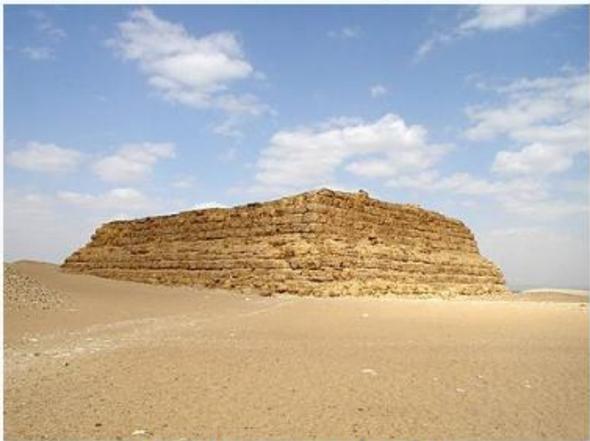

*Screenshot from images Courtesy Wikipedia. The picture of the mastaba is a courtesy Jon Bodsworth - www.egyptarchive.co.uk . "Instead of the last sign, which depicts a pyramid, a similar sign without a point signifying 'mastaba' must have actually been used, which cannot be shown for technical reasons." Ref.[1] is Gundacker, 2009.*

The mastaba of mud bricks is also shown in https://egyptphoto.ncf.ca; "The riddle of why Shepseskaf built his tomb here, and why he chooses a mastaba like structure is not yet resolved by Egyptologist. Certainly, it has something to do with the transition from the 4th to the 5th dynasties." No reference given. "Certainly", why "certainly"?

https://www.britannica.com/place/Saqqarah#ref633419; "Shepseskaf of the 4th dynasty (c. 2575–c. 2465 bce) built Maṣṭabat Firʿawn, a coffin-shaped tomb, and several kings of the 5th dynasty (c. 2465–c. 2325 bce) also constructed their pyramids at Ṣaqqārah. Unas, the last king of the 5th dynasty, was the first to inscribe on the walls of his pyramid chambers the Pyramid Texts, … Succeeding kings of the 6th dynasty (c. 2325–c. 2150 bce) continued the practice of inscribing Pyramid Texts in the underground chambers. With the exception of Teti, the 6th-dynasty kings built their pyramids to the south of Unas's pyramid, and the most southerly is that of a 13th-dynasty (c. 1756–c. 1630 bce) king".

**The Userkaf tomb (pyramid)**

"The pyramid complex of Userkaf was built c. 2490 BC[1] for the king Userkaf, founder of the 5th Dynasty of Egypt (c. 2494–2345 BC). It is located in the pyramid field at Saqqara, on the north-east of the step pyramid of Djoser. Constructed in dressed stone with a core of rubble, the pyramid is now ruined and resembles a conical hill in the sands of Saqqara.[1] For this reason, it is known locally as El-Haram el-Maharbish, the "Heap of Stone",[2] and was recognized as a royal pyramid by western archaeologists in the 19th century" (Wikipedia, mentioning [1] Lehner, 1997, and [2] Lauer, 1988).

"The complex is markedly different from those built during the 4th Dynasty (c. 2613–2494 BC) in its size, architecture and location, being at Saqqara rather than the Giza Plateau. As such, Userkaf's pyramid complex could be a manifestation of the profound changes in the ideology of kingship that took place between the 4th and 5th dynasties,[1] changes that may have started during the reign of Userkaf's likely immediate predecessor, Shepseskaf.[5]" (Wikipedia, mentioning [1] Lehner, 1997, and [5] Shaw (Ed.), 2000).

**Dating the pyramids**

In Magli's arXiv, we find: "Unfortunately, besides a controversial proposal (Gautschy et al. 2017) the first available registrations come from the Middle Kingdom, so no Sothic anchor is as yet available for the Old Kingdom. *However*, astronomy in Egypt can be measured also in architecture. In particular, *strong hints* pointing to the planning of the Khufu pyramid – **and thus to Khufu's first year - in 2550 bC. with the impressive uncertainty of only 10 years** come from the alignment of this pyramid to the cardinal points" (Magli mentioning Belmonte, 2001).

Magli does not propose dates for the other kings from Belmonte. For further references, see Puchkov, 2023.

In Belmonte, 2001, we find the following, from https://adsabs.harvard.edu/pdf/2001JHAS...32....1B.

"According to my [Belmonte] proposal, the pyramid of Khufu would have been aligned between 2571 and 2565 BC, at a time when Phecda and Megrez were at upper culmination, since, … with the pair of stars at lower culmination, the dates (2559 to 2553 BC) do not fit the chronological pattern accuracy. *If we assume, following Spence, that the alignment was carried out in the first year of the king's reign,* we obtain *dates for Khufu's ascension to the throne* that are just in the middle of the highest and lowest chronologies accepted today by Egyptologists (see Table 1 [in Belmonte], column 3 and 4" (see the further discussion in Belmonte, 2001). Spence is Spence, 2000.

In the Table 1 from Belmonte, 2001, we have the dates of other kings. Please consider the following Table adapted from a screenshot of the Belmonte's work. We report only Khufu's chronology.

| King | Regnal Years | High Chronology | Low Chronology | Astronomical Alignment Dates | Dates New Proposal |
|---|---|---|---|---|---|
| Khufu | 23 | 2589-2566 | 2551-2528 | 2571 to 2565 | 2577-2554 |

Note the new proposal of dates by Belmonte. The high chronology is from Malek, and the low chronology from Baines and Malek. According to Table 1 from Belmonte, the reign of Khufu lasted from 2577 BC to 2554 BC. The first year was 2577 BC.

We can find in Belmonte's text the note 30, where the author mentions Malek, Shaw, Clayton and Baines and Malek.

Tupikova, I. (2022) explains: "Belmonte based his method upon the same assumption as K. Spence: that the real goal of the pyramid orientation was the North Celestial Pole. According to this author, Khufu's pyramid could have been aligned between 2571 and 2565 BC, at the time when Phecda and Megrez were at their upper vertical alignment; that gives "*dates for Khufu's ascension to the throne that are just in the middle of the highest and lowest chronologies accepted today by Egyptologists*" (Belmonte 2001, S12). In note [55], Tupikova tells: "The author considers in his text the high chronology according to J. Malek (Shaw 2000, 89–117) who assumes for Khufu's reign 2589–2566 BC and low chronology according to Baines and Malek (1980, 36) with the dates of 2551-2528 BC. Belmonte (2001, S11) correctly rejected the vertical alignment of Phecda and Megrez at lower culmination, because "the pair of stars at lower culmination, …do not fit the chronological pattern accurately". "With the actual precession theory, we can state that for the location of Khufu's pyramid, the vertical through Phecda and Megrez crossed the meridian around 2552–2551 BC" (Tupikova, adding in note, that a " more accurate date cannot be calculated because of the absence of a nutation calculation for the time in question; according to our estimation, it might change the date of the vertical alignment in the range of ± 1 year").

Here the abstract of Tupikova, 2022. "The remarkable degree of accuracy with which some of the Old Kingdom pyramids are oriented towards the cardinal directions is one of the most challenging problems in the history of science. The progressive deviation of the orientation of the 4th Dynasty pyramids from true north was long understood to be a consequence of the pyramids having been aligned to a star whose celestial position changed due to the effect of the general precession of the rotational axis of the Earth. Instead of a single star, recent proposals considered a possible orientation towards some notable vertical or horizontal stellar configurations. The main idea behind these recent attempts at explanation was to justify the gradual deviation of the pyramid alignments by way of the selected target stellar configuration exhibiting a similar azimuthal trend. Considering conventional Egyptian chronologies of this period to be only relative, and the astronomically determined data to be fully reliable, the researchers tried to make the two trends match perfectly by shifting the conventional Old Kingdom chronologies by some, often significant, number of years. Too little attention, however, was paid to allowing for systematic and random errors in the surveying of stars and in the orientation of pyramids towards the observed asterism, which may obfuscate the real accuracy of the methods and conceal the actual targets of observations. In this text, we consider recent proposals and analyze their errors. We propose and discuss two new solutions whose systematic errors are minimal among all the known proposals: one based upon the horizontal alignment of Alioth and Mizar, and another one upon the vertical alignment of Kochab and zeta-UMi. In contrast to other methods, the latter pair has the advantage that it could have been observed at lower altitudes. Both variants show an impressive degree of agreement with the trend in the orientation of the pyramids for von Beckerath's (lower estimates) as well as for Baines and Malek's chronologies of the period. It appears to us that the preserved Egyptian astronomical diagrams are fully consistent with our new proposals" (Tupikova, 2022).

Here is a chronological Table from Tupikova.

"Due to the effects of precession, stellar azimuths are time-dependent and can only be calculated for specified calendar dates. Several chronologies of the period are available; in this text we have used the four most agreed-upon chronologies—von Beckerath's (1997), Baines and Malek's (1980), Malek's (as given in Shaw 2000) and Hornung et al. (2006), all modified according to Stadelmann's proposal by having 48-years for the duration of Snofru's reign. The corresponding data in years BC are given in Table 2 where the dates in square brackets for the construction of Snofru's Bent and Red pyramids follow the temporal proportions given in Stadelmann (1986)" (Tupikova, 2022).

| Ruler/pyramid | Dynasty | Acession date (Shaw) | Acession date (Baines&Malek) | Acession date (von Beckerath) | Acession date (Hornung et al) |
|---|---|---|---|---|---|
| Snofru-Meidum | 4 | 2637 | 2599 | 2602 | 2558 |
| Snofru-Bent | 4 | [2620] | [2580] | [2583] | [2541] |
| Snofru-Red | 4 | [2609] | [2569] | [2572] | [2530] |
| Khufu | 4 | 2589 | 2551 | 2554 | 2509 |
| Djedefre | 4 | 2566 | 2528 | 2530 | 2482 |
| Khafre | 4 | 2558 | 2520 | 2522 | 2472 |
| Menkaure | 4 | 2532 | 2490 | 2489 | 2447 |
| Sahure | 5 | 2487 | 2458 | 2446 | 2428 |
| Neferirkare | 5 | 2475 | 2446 | 2433 | 2415 |
| Unas | 5 | 2375 | 2356 | 2317 | 2321 |

Table 2 in Tupikova, 2022.

In Tupikova we find mentioned Puchkov. "Puchkov also refers to the chronology proposed by J. Breasted (1906 1907, I, 40–47) who suggested dates about 280 years earlier for the 4th Dynasty. The starting dates for the alignment ceremonies used by A. Puchkov (col. 3) for the synchronization of the relative and absolute (astronomical) chronologies together with his new proposed dates (col. 4) are given in Table 3; the author gives ± 5 years as a confidence interval" (Tupikova, 2022).

| Ruler/pyramid | Dynasty | Date of alignment | Date of alignment (Puchkov) |
|---|---|---|---|
| Djoser | 3 | 2666 | 2888 |
| Snofru-Meidum | 4 | 2636 | 2858 |
| Snofru-Bent | 4 | 2619 | 2841 |
| Khufu | 4 | 2588 | 2810 |
| Djedefre | 4 | 2565 | 2787 |
| Khafre | 4 | 2557 | ?(2779) |
| Menkaure | 4 | 2531 | 2753 |

Table 3 from Tupikova, 2022.

In Spence, 2000, Table 1 reports the following dates ("The dates in column 2 are from von Beckerath's chronology (**lower estimates**)8"). Ref.8 is Beckerath, Chronologie des pharonischen Aegypten, 1997.

| Khufu | Khafre | Menkaure | Sahure | Neferirkare | Unas |
|---|---|---|---|---|---|
| 2554 BC | 2522 BC | 2489 BC | 2446 BC | 2433 BC | 2317 BC |

To have the higher estimate, here von Beckerath's dates from Gautschy et al., 2017:

| Khufu | Unas |
|---|---|
| 2604/2554 BCE | 2367/2317 BCE |

In Gautschy et al., 2017, we have data from Shaw, 2000, given as:

| Khufu | Unas |
|---|---|
| 2589 BCE | 2375 BCE |

Here Shaw, 2000, page 483 (*note the gap of four years between the fourth and the fifth dynasties*).

| 4th Dynasty | 2613–2494 | 5th Dynasty | 2494–2345 |
|---|---|---|---|
| Sneferu | 2613–2589 | Userkaf | 2494–2487 |
| Khufu (Cheops) | 2589–2566 | Sahura | 2487–2475 |
| Djedefra (Radjedef) | 2566–2558 | Neferirkara | 2475–2455 |
| Khafra (Chephren) | 2558–2532 | Shepseskara | 2455–2448 |
| Menkaura (Mycerinus) | 2532–2503 | Raneferef | 2448–2445 |
| Shepseskaf | 2503–2498 | Nyuserra | 2445–2421 |
| | | Menkauhor | 2421–2414 |
| | | Djedkara | 2414–2375 |
| | | Unas | 2375–2345 |

Regarding Khufu, Robinson, 2022. "Without a doubt, the Great Pyramid was commissioned by the Old Kingdom pharaoh Khufu (Cheops). The British Museum and Cairo's Egyptian Museum give his regnal dates as 2589 to 2566 BCE [that is, Shaw, 2000]. *Egyptologists Mark Lehner, who has conducted fieldwork at Giza for four decades, and Zahi Hawass, a former Egyptian government official in charge of Giza, argued for the later range of 2509 to 2483 BCE in their massive 2017 book, Giza and the Pyramids.* But another high-profile Egyptologist, Pierre Tallet, whose pioneering fieldwork on the Red Sea coast of Egypt began in 2011, favors the earlier range of 2633 to 2605 BCE, derived from a recent astronomically based chronological model for the Old Kingdom" (Robinson, 2022).

In Morishima et al., 2017, we find told: "The Great Pyramid or Khufu's Pyramid was built on the Giza Plateau (Egypt) during the IVth dynasty by the pharaoh Khufu (Cheops), who reigned from 2509 to 2483 BC", and the reference is Lehner, 2008. The first year, 2509, tells us that Lehner used Hornung et al.. However, in Lehner, 2007, the author tells that the "dates used here are based on the chronology developed by Professor John Baines and Dr Jaromir Maiek and set out in their Atlas of Ancient Egypt".

4$^{th}$ dynasty (Lehner, 2007)

| Sneferu | Khufu | Djedefre | Khafre | Menkaure | Shepsekaf |
|---|---|---|---|---|---|
| 2575-2551 | 2551-2528 | 2528-2520 | 2520-2494 | 2490-2472 | 2472-2467 |

The 5$^{th}$ dynasty is given from 2465 to 2323, with Userkaf reigning from 2465 to 2458 BC.

We have therefore that Lehner changed his mind. The reign of Khufu passed from 2551-2528 to 2509-2483. The first regnal year shifts of 42 years. We can expect a corresponding shift for Shepsekaf, passing from 2472 to about 2430. In this case, the total eclipse of 2471 BC was before his reign. It happened during the reign of Khafre. If we consider the uncertainty of ±25 years, the same that Magli assumed, we can conclude that, using Hornung et al. chronology, Shepsekaf had not observed the eclipse.

**Sellers' Table**

Regarding the eclipse of 2471 BC, Magli is mentioning Sellers, 1992, in the following manner: "So, we are led to search if a total solar eclipse occurred on Lower Egypt in a date compatible with Shepseskaf first year. This eclipse actually exists and occurred in the morning of April 1, 2471 b.C. (this fact has been noticed before but without deepen the investigation by Sellers (1992) and Magli (2013))."

The book by Sellers is available thanks to archive.org. Here some passages.

Page 132: "*In 2471 BC a total solar eclipse is predicted to have taken place over the sacred city of Pe, north on the Delta*." "About 2473 BC we find sun temples built some miles to the South of Giza. Six rulers in the Fifth Dynasty built these great temples to the sun and named them such names as Pleasure of Re, Horizon of Re and Field of Re. Each had a small obelisk perched on a square base like a truncated pyramid and recalled the very ancient stone at Heliopolis known as the benben. The phrase, 'the radiant one', derives from this word. In 2471 BC a total solar eclipse is predicted to have taken place over the sacred city of Pe, north od the Delta. At the beginning of the Sixth Dynasty another change took place. The religious fervor honoring the sun god which had marked almost all the effort of the Fifth Dynasty, shifts its emphasis to the worship of Osiris" (Sellers, 1992).

Page 277: Pe 1 April 2471 Userkaf c. 2480 (Dyn. 5).

***Consider that Pe is also known as Buto.***

The page proposes a table. Table caption: Conjectured accession dates and all predicted total eclipses over centers of power in Ancient Egypt.

Here we repropose five of the oldest eclipses.

| City | Date (BC) | King | Conjectures accession date (BC) |
|---|---|---|---|
| Nekhen | 5 Feb 3109 | Menes | c. 3110 (Dyn. I – Parker) |
| Thinis | 28 Feb 3046 | Aha | c. 3048 (Dyn. I – Manetho) |
| **Pe** | **1 April 2471** | **Userkaf** | **c. 2480 (Dyn. 5 – Gardiner)** |
| Nekhen | 23 March 2340 | Teti | c. 2341 (Dyn. 6 - Parker) |
| Memphis | 29 June 2159 | Akhtoy | c. 2154 (Dyn.9 - Parker) |

Page 286: "On 1 April 2471 BC at 7:23 a.m., with the sun large on the horizon, a total eclipse is predicted to have been witnessed by the inhabitants of the sacred city of Pe, a city North in the Delta. At Memphis and Heliopolis this eclipse of 2471 BC would have been seen as near total".

"Sir ***Alan Gardiner gives 2480 BC as the beginning of the Fifth Dynasty***, a dynasty distinguishable by its invention of a new type of monument, a square obelisk on a square base, fronted by a raise terrace with a large alabaster altar. The sun god could now be worshipped under an open sky. North of this altar was a large area where it appears that the slaughter of bulls took place. A long covered passage, decorated with beautiful scenes, some representing the Sed Festival, led to the obelisk platform" (Sellers, 1992).

In Sellers' book, we do not find Shepseskaf, but we find Userkaf.

In Kelley and Milone, 2011, we find also Userkaf. See the Table 8.3 (Egyptian dynastic eclipses) in the book.

| Chronology Gardiner | Chronology Clayton | Egyptian capitals | Total solar eclipse dates over capitals |
|---|---|---|---|
| V - 2480 BC | V – Userkaf 2498 BC | Pe | 2471 Apr. 1, BC |

Magli, 2024: "This eclipse actually exists and occurred in the morning of April 1, 2471 b.C. (this fact has been noticed before but *without deepen the investigation by Sellers* (1992) and Magli (2013))."

***How deep was the investigation made by Sellers?*** Here, in the following the discussion in Kelley and Milone, 2011: "Sellers also found some suggestive evidence that dynastic changes in Egypt sometimes occurred following an eclipse that was total at the capital of the particular dynasty that lost power. Given uncertainties in the back-calculation of eclipses and other uncertainties in the calculation of ancient Egyptian chronology, it would be rash to consider this more than a reasonable hypothesis, but given the identification of the Pharaoh as the son of the Sun, one would expect solar eclipses to be regarded as extremely bad omens. The coincidences of dynastic changes with eclipse dates as calculated by Sellers, shown in Table 8.3, are more than we would have expected. We should make it clear that Sellers is promising a causal connection based in the emotional reactions of the Egyptian people in terms of their mythology. If there was such a correlation, later people might have interpreted it as astrological causation, but nothing indicates that this was an idea present in Egypt earlier than, perhaps, the time of the Assyrian invasions" (Kelley and Milone, 2011.

In Steele, 2013, it is told that "we have very little astronomical material from Egypt before the Hellenistic period. Most of what we have relates to time-keeping and the calendar. It has been claimed by Sellers (1992), although without fully convincing evidence, that eclipses played a fundamental role in development of Egyptian religion, but no actual records of eclipses have so far been found from ancient Egypt".

In fact, Sellers, 1992, was the first researcher that linked eclipses to dynastic changes.

***Where is the causal connection?*** The fall of a dynasty can have several tangible reasons, besides the bad omen of an eclipse.

Or, for instance, a king marries an influential princess of the preceding dynasty.

"Neferhetepes is the only know wife of Userkaf, and was mother of his son and successor, Sahura, … Neferhetepes had been considered the daughter of Djedefra of the Fourth Dynasty by some scholars …, but other have pointed out that this would give her an impossible long life span … . With the new evidence from Abusir, her historical position is clear but not necessarily who her parents were. Was Neferhetepes a daughter of Menkaura and married to Userkaf to help secure his right to the throne?" (Sabbahy, 2020, mentioning El-Awady, 2006).

Regarding eclipses observed from Egypt, in Belmonte and Lull, 2023, page 518, it is told that the "absence of sources is especially dramatic", "especially the total ones". "This apparent absence of information … has puzzled many historians and scientists" (Belmonte and Lull do not provide a list of references about the many historians and scientists). "However, according to some researchers, the appearance of the sun during the phase of totality was preserved in a symbolic form". Belmonte and Lull do not provide a list of references regarding the Egyptian representative symbolic forms of an eclipse, but they continue telling "for example, Jane Sellers, 1992, has proposed that part of the myth of the struggle between Seth and Horus could be reinterpreted in light of a phenomenology related to a solar eclipse, *although **her ideas are very controversial** and have **not been widely accepted**"*. Belmonte and Lull do not explain why "her ideas are very controversial". Once more, the burden of explaining which of the Sellers' ideas are controversial is with Belmonte and Lull. Is the Sellers' link between Seth and Horus myth with eclipse controversial or not?

"Another idea proposed by Sellers (1992) is that total solar eclipses may have had something to do with certain significant dynastic changes in ancient Egypt, especially those of a dramatic nature. This idea has been furthest developed in the Internet by a series of researchers such as Leo Duval in France, the American William McMurray and The Egyptian Ahmed Ibrahim" (Belmonte and Lull, 2023). Again, we find told by Belmonte and Lull that the a "controversial idea" exists about erection of obelisks and eclipses.

In Ruiz, 2001, we find the legend of Apep/Apophis, personification of the darkness of the night. No references given by Ruiz. "During storms that darkened the sun. and at time of eclipse, the Egyptian believed that Apep was getting the upper hand. One can imagine the fear of the deeply spiritual and superstitious Egyptians during a total solar eclipse" (Ruiz, 2001).

**The magnitude of an eclipse**

Smith, 2012, part 2: "Research has suggested that eclipses at a particular place tend to occur in clusters especially if they occur at sunrise or sunset. Such eclipses, as will be seen later, may allow the obscuration of the solar disk to be much more obvious and therefore more likely than normal to have been noticed. Brewer (1991, p.70) illustrates the flavour of the apparently random nature of the recurrence interval by quoting a table of examples ranging from 837 years for London (from 29 Oct 878 AD to 22 Apr 1715 AD) to 1½ years in Southern New Guinea (from 11 Jun 1983 AD to 22 Nov 1984 AD), listing two for places in Egypt – 312 years for Giza (1 Apr 2471 BC to 29 June 2159 BC) and 9 years for Thebes (31 May 957 BC to 22 May 948 BC). Although total eclipses are rare, near total eclipses are more common but can be almost as dramatic. The precise degree of darkness achieved during a solar eclipse will depend on many factors, including the time of day, weather conditions and the cloudiness of the sky, but the major factor in determining this will be how much of the solar disk is covered by the moon. Astronomers call this the magnitude of the eclipse, measured simply by the linear fraction of the solar disk obscured by the moon … . In practice this magnitude (μ) can range from 0, when the moon's disk is about to touch the sun's disk, up to 1.08, when the moon's disk appears slightly larger than the sun."

Smith, 2012, part 1: "Solar eclipses in Egypt have been discussed by several researchers in recent years. *Sellers* (1992) considered that there may have been a link between such events and Pharaonic accession, while *Ibrahem* studied possible correlations of solar eclipses with key events or inscriptions, although the eclipse predictions he used are no longer accurate. *Aubourg* (1995), using the motions of the planets to study the dating of the Zodiac of Dendera, … . This research, however, has been disputed recently and a different dating … *McMurray* (2003 & 2004), using the latest predictions, notes the possible influence of a solar eclipse on Akhenaten and has also been attempting to correlate lunar and solar eclipse dates with dateable inscriptions to try to develop an absolute chronology. *Ryholt (2011)* has concluded that the association with astrology of Necho II (who gained the epithet 'the wise') may have been due to an historical eclipse marking the beginning of his reign – the same eclipse mentioned in the Neue demotische Erzählung but which was associated there with the earlier death of Necho II's predecessor Psammetichus. However, more recently *Park* (2012) has argued that Ryholt's analysis does not put forward sufficient evidence … Although this [Smith's] paper concerns work on eclipses in Ancient Egypt, it is worth noting in passing the related work on the Asiatic Campaigning of Horemheb (*Redford* 1973) which has implications for Egyptian chronology because of the political relationship and interaction between the Egyptian and the Hittite empires. Redford draws attention to the annals of Mursilis II's tenth year which mention an "omen of the sun", generally accepted as a solar eclipse. Drawing upon his earlier work in this area, he argues that this is likely to have been the eclipse of March 13th 1335 BC and proceeds to use this date to determine an absolute chronology of Thutmosis III to Horemheb. Recent research has suggested that this date is too early and that there are other more likely candidates for this eclipse. The first is that of 24th June 1312 BC, which passed over Northern Anatolia close to Hattusa where Mursilis II and his men were likely to have been based (*Bryce* 1998). An alternative view has been put forward that the eclipse was instead that of 13th April 1308 BC (*Åström*, 1993)" (Smith, 2012).

Smith is mentioning Gautschy, 2012.

**ΔT and its uncertainty**

Gautschy, 2012: "For many centuries, the fundamental unit of time was the rotational period of the Earth with respect to the Sun. Universal Time (UT), also called Greenwich Mean Time (GMT), is based on the mean solar time at Greenwich. Unfortunately, UT is not a uniform time scale over historical times because Earth's rotational period gradually decreases. Therefore, the calculation of local circumstances of solar eclipses in the far past is subject to uncertainties. As the Earth rotates, tidal friction, inflicted by the gravitational attraction of the Moon and the Sun is at work. The result is a transfer of angular momentum from the Earth to the Moon. The Earth loses energy and slows down, the Moon gains energy and its distance from the Earth increases. Today, atomic clocks are used for accurate timekeeping. Terrestrial Dynamical Time (TDT) is such an atomic time scale. Solar eclipse calculations are based on TDT, the position of the visibility area of an eclipse depends however on UT. Calculations from TDT have to be converted into UT; therefore the time difference between TDT and UT must be known. This time difference, called ΔT, which adds up to about 12 hours around 2000 BC and the uncertainty of ΔT - about 2 hours in 2000 BC - has to be taken into account in the calculations. For more information about ΔT see e.g. the webpage of Robert van Gent" (Gautschy. 2012). https://www.gautschy.ch/~rita/archast/solec/solec.html

**In a newsletter**

We can find a proposal from Andis Kaulins <AKaulins@AOL.COM>, Tuesday, July 10, 2001 2:59 PM Subject: Solar Eclipses - Absolute Chronology of Egypt. The title is The Absolute Chronology of the Pharaohs by Solar Eclipses in Egypt", eclipse.gsfc.nasa.gov

"The first original hypothesis presented [by Kaulins] is that Pharaohs put their pyramids and tombs on the paths of solar eclipses and that this can be calculated to a day. A second original hypothesis is that the causeways to the pyramids may mark the path of question. A third hypothesis is that these eclipses are "archived" in the hieroglyphs of the Pharaohs as their "name" - e.g. the "shaded" or "eclipsed" object in the hieroglyph of Cheops. Fourth, there is the original hypothesis that the RA or "Sun Name" of a pharaoh in the cartouche gives the position of the Sun in the heavens and the "AMUN-name" of a pharaoh shows the position of the Moon. Possibly, the analysis below may also permit astronomers to get an absolutely accurate T-Delta value regarding the change in the rate of the spin of the earth in the last 5000 years (there is dispute about this value among astronomers). To obtain the correct value, I would suggest a combination of approaches to get a properly calibrated Delta T value  …" (Kaulins, 2001). See further discussion in the Newsletter.

https://eclipse.gsfc.nasa.gov/SENL/SENL200108A.pdf

"The idea of eclipses as being of such great importance came from a posting to me by ChasInca on Shalmaneser (Solomon-Nezer) and from the internet page of Amir Bey … who used the following sources -Solar Eclipses of the Ancient Near East, by M. Kudlek and E.H. Mickler. ---The Canon of Solar Eclipses, by J. Meeus and H. Mucke -Solar Fire 4.07, a software program developed by Esoteric

Technologies -Lunar Eclipses of the Ancient Near East by M. Kudlek and E. H. Mickler Bey's data of course is subject to verification but - he provides a great graph of the path of the eclipses - and I think that his data will in general be confirmed" (Kaulins, 2001).

The Amir Bey discussion is in [web.archive.org](web.archive.org).

"April 1, 2471 BC at 9 a.m. - Solar Eclipse over Egypt at the cross of the ecliptic and the celestial equator at the Pleiades Aldebaran. Since Kate Spence, Egyptologist at Cambridge, recently put the building of the Great Pyramid to 2467 BC plus or minus 5 years, I had regarded this to have been the event which marked the begin of the Cheops pyramid. However, this was wrong. The April 1, 2471 BC eclipse applied to the "Red Pyramid" (for the red tar Aldebaran) and more northerly of the two pyramids of Snofru" (Kaulins, 2001).

This is the screenshot from Kaulins, regarding the eclipse of 2471 BC

The newsletter was open to comment. And then we find the following from R.H. van Gent r.h.vangent@PHYS.UU.NL. Van Gent criticized Kaulins.

R. van Gent observes: "The recent attempt of Andis Kaulins to re-date the ancient history of Egypt by means of solar eclipses provides a nice example of the various pitfalls one can fall into. … Andis Kaulins is probably unaware that he is here following a process that in the past Robert R. Newton has aptly named the 'eclipse game'. These are the rules:

a) Find an ambiguously stated historical record and convince yourself that it can only refer to a total solar eclipse and nothing else.

b) Make sure that the record cannot be dated chronologically within the nearest half century or so.

c) Find the nearest solar eclipse in time from a calculated list of eclipses that satisfies your interpretation of the record. Preferably pick an eclipse list that is not too specific as to how it was calculated.

d) Date the record with the help of this eclipse and convince historians that the record has now been dated firmly with the help of precise astronomical techniques.

e) Some years later, astronomers will stumble across this record in a historical publication (probably several removes away from the original publication), believe that it has been dated firmly by historical means and determine an improved value of Delta T for that epoch. This Delta T value can then be used to calculate 'improved' eclipse tables for the aid of historians."(Van Gent, 2001)

"It cannot be stressed too much that historical dates inferred in this way are completely meaningless unless they can be corroborated by other means." (Van Gent, 2001).

"Concluding, I cannot say that I am in the least convinced by Andis Kaulin's astronomical re -dating of the ancient history of Egypt by means of his suggested eclipse records. For me his theory already fails in step (a) of the above mentioned 'eclipse game'. I see no reason why every ancient Sun symbol (or even round symbol) should have to refer to a solar eclipse. There may certainly be hidden eclipse records in Egyptian texts and inscriptions but I do not believe them to be found in the names and cartouches of the Egyptian pharaohs. Sun symbols can mean a lot of things, not all necessarily related to astronomy. The University of Utrecht also uses a solar image in its logo but there is definitely no link with any solar eclipse visible from Utrecht around the date of its founding (though I am sure there will be people in a far distant future who will be tempted to interpret it as such)" (Van Gent, 2001).

**Other eclipses**

In Kaulins, we find two years after another eclipse, September 2, 2469 BC. Kaulins links this eclipse to the Snefru pyramids, in particular to the building of the second pyramids. "Solar Eclipse at 9 a.m. at Spica in Virgo. This led to the building of the second of the two pyramids of Snofru and the reason that a second one was built was this second eclipse - only 1 & 1/2 years apart - so close in time to the first. It is now easy to determine which of the two Snofru pyramids was built first - about which the scholars have long disputed - the center of the 2469 eclipse was further south, and this is the "Bent Pyramid"." (Kaulins)

We can use the web site http://ytliu.epizy.com/eclipse/solar_local.html?i=1; selecting Giza, Egypt (many thanks to the person that suggested me this site, thank you).

Here the screenshot of eclipses, close to that of 2471 BC (that is, -2460)

Display time in: ○ specified time zone (UT1+2)   ● local apparent solar time

| Calendar Date | Ecl. Type | Partial Eclipse Begins | Sun Alt (°) | A or T Eclipse Begins | Maximum Eclipse | Sun Alt (°) | Sun Azm (°) | A or T Eclipse Ends | Partial Eclipse Ends | Sun Alt (°) | Ecl. Mag. | Ecl. Obsc. | A or T Ecl. Dur. |
|---|---|---|---|---|---|---|---|---|---|---|---|---|---|
| -2472-Nov-15 | P | 16:00:32 | 18 | — | 17:11:35 | 3 | 252 | — | 17:29.7(s) | 0 | 0.834 | 0.783 | — |
| -2470-Apr-01 | P | 07:01:59 | 11 | — | 08:06:11 | 25 | 111 | — | 09:17:36 | 38 | 0.989 | 0.993 | — |
| -2468-Sep-02 | P | 17:11:44 | 18 | — | 18:08:59 | 6 | 284 | — | 18:37.5(s) | 0 | 0.936 | 0.923 | — |
| -2463-Nov-06 | P | 11:19:43 | 49 | — | 12:53:46 | 48 | 200 | — | 14:26:37 | 37 | 0.942 | 0.913 | — |

The following links are the simulations: 2473 BC, 2471 BC, 2469 BC, 2464 BC .

**Buto**

The eclipse, that of 2471 BC, was "dramatic" in Buto, according to Magli. Why in Buto and not in another place? "Buto was an important cult centre, located in the delta. The Cobra-goddess Wadjet was worshipped there" (Magli, 2024). Magli does not provide references about the importance of Buto and the goddess worshipped there.

We have told before that the Shepseskaf's mastaba was like a huge coffin. In Magli, 2024: "Another, more reasonable interpretation is that the tomb resembles a specific, sacred building: a "Buto shrine" (Lehner 1999)". Why, this interpretation is more reasonable? Is Lehner proposing this interpretation, that is, that the Shepseskaf's mastaba represents the Buto shrine. Or did Lehner just describe the Buto Shrine? Is the link between Shepseskaf's mastaba and Buto shrine a Magli's speculation?

Searching in Google Books, we can propose a sentence in a Magli's book, 2013:

"This leads us to the second problem, that of understanding the shape, which is not precisely a bench" but has two vaulted ends. It is usually said in the literature that the monument resembles a "Buto shrine"; however, no explanation has been given for such a strange choice of form (Lehner 1999).

The city of Buto was a sacred centre, …", where Lehrer is mentioned. In Magli, 2024, the author writes that: "*Information about Shepseskaf are [is] scarce*: we have, however, his tomb (Jéquier 1925)".

However, we can find a detailed answer in https://www.touregypt.net/, where we find "speculation" (https://archive.is/bVRsk, 2024, https://archive.is/ibFP , 2012, in both pages we find mentioned the Buto shrine).

"The aberration of Shepseskaf's name, his tomb and the tomb of his possible daughter, consort or/and half-sister all stand out like sore thumbs, awaiting the theories of Egyptologists that may perhaps never be proven. All we can do here is present the current speculation, and possibly add a little of our own. Jéquier offers an initial explanation that other Egyptologists, such as Jaromir Malek, who provided the Old Kingdom component of the Oxford History of Ancient Egypt, find tempting. He was rather convinced that Shepseskaf choose the mastaba style tomb as an *intentional protest against the priesthood of the cult of Re,* the sun god, which was gaining considerable influence. Jéquier believed that the ancient Egyptians considered the pyramid a symbol of the sun, as do many modern Egyptologists. Certainly, the rise of the pyramid coincided with the growing influence of Re's cult. He also believed that Shepseskaf's move away from the Giza Plateau and hence, the traditions of his immediate predecessors, supported his position, but perhaps even more important to his argument was Shepseskaf's abandonment of Re's reference within his name." (touregypt.net)

"This theory, along with several of its components, can be easily attacked and has been from a number of different directions. One of the easiest elements to overcome in Jequier's theory is Shepseskaf's move away from the Giza Plateau. *His father, Menkaure was required, due to spatial restrictions, to place his pyramid far away from the Nile,* and it is relatively clear from his valley temple placement, blocking the principal conduit for construction materials into the necropolis, that he intended no more major monuments to be built there. *In fact, there was simply no more room for such a major construct on the Plateau. This undoubtedly prompted Shepseskaf to look for another location*, and in doing so, he chose a place that not so very far from the pyramids of the dynasty's founders. In fact, the stone for his mastaba came from Dahshur, the location of Snefru's Bent and Red Pyramids. Saqqara was also a very ancient necropolis, that in fact relates somewhat to his use of a mastaba rather than a pyramid" (touregypt.net)

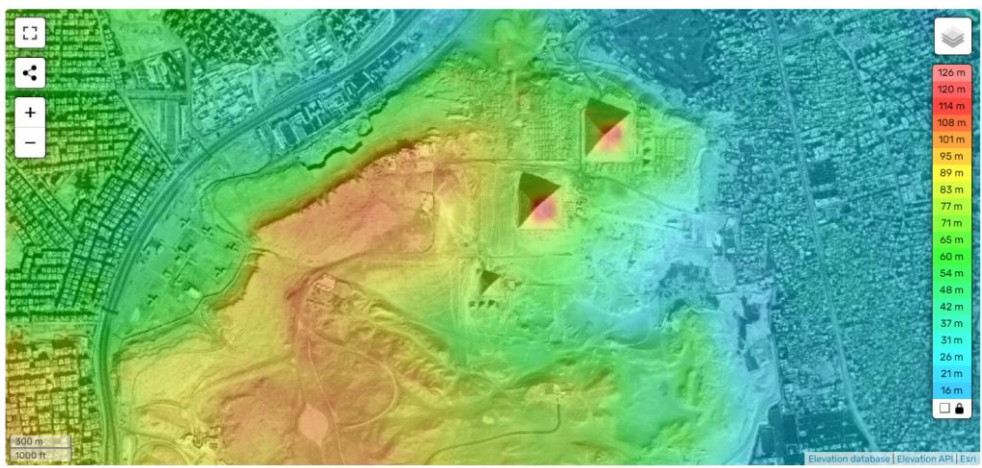

Topographic map courtesy TessaDEM: https://it-it.topographic-map.com/map-2f2frr/الجيزة/

"Regarding Shepseskaf's use of a mastaba rather than a pyramid as a protest against the priesthood of Re, Ricke believed that the obelisk, rather than the pyramid, was considered by the Egyptians to be the symbol of the sun. After all, the 5th Dynasty kings who we believe constructed the sun temples, mostly at Abu Ghurob, with a short obelisk as a focal point, did so in addition to their pyramid complexes mostly at Abusir. In his opinion, which seems to be mirrored by one of modern Egypt's great scholars, *Mark Lehner, he was, rather than rejecting the cult of re, honoring his religious heritage in the form of the Lower Egyptian "Buto-type" tomb.* It was really not very uncommon at all for Egyptian pharaohs to display such archaic tastes. Similarly, Hans-Wolfgang Muller (1907-1991) felt that Shepseskaf's mastaba was *a huge version of a hut hung with matting*. Indeed, Stadelmann, drawing on the arguments of Ricke and Muller, pointed out that Shepseskaf's use of niches in the courtyard of his mortuary temple, as well as in certain elements of his father's pyramid complex, was, an archaizing element from Egypt's earliest architecture" (touregypt.net)

That is, Lehner proposed a "Buto-type" tomb.

"In addition*, it must also be noted that Shepseskaf faced the difficult task of completing his father's pyramid at Giza. This must have certainly created a considerable administrative and financial burden, at a time when the Egypt was apparently suffering some economic hardship.* This may have led him to downsize his own tomb. Other possibilities exist. It is possible that the mastaba was initiated prior to his ascent to the throne, for example, or that it was a provisional tomb created with the possibility that if time permitted, another once could have been built" (touregypt.net)

"We question whether many of the issues will ever be answered…. But the possibility always exists that future discoveries may, at least, provide answers to at least some of the questions surrounding this mysterious man and his tomb" (touregypt.net). Eclipse game?

The website gives references: Clayton, 1994, Lehner, 1997, Grimal, 1988, Dodson, 1995, Shaw, 2000, Verner, 2001.

In the version archived in 2012: "Shepseskaf's mastabas was huge, measuring some 99.6 meters (327 ft) long by 74.4 meters (244 ft) broad, and oriented north to south. The core of the mastaba was built in two levels of large, grayish yellow limestone blocks that originated in the stone quarries west of the pyramids at Dahshur. In the early years of Egyptian exploration, it was still possible to find remnants of the pathways over which this stone was transported. The mastaba was encased with fine white limestone except for the very bottom course of red granite (which makes us wonder if it was left over from his father's complex). On some of the casing blocks may be found inscriptions of Prince Khaemwese's later restoration of this monument. The outer slope of the casing was 70º and *it had a vaulted top between vertical ends, taking the shape of a Buto shrine (according to some Egyptologists, such as Mark Lehner)*" (touregypt.net).

**Pe and the Buto shrine**

https://www.treccani.it/enciclopedia/buto_(Dizionario-di-Storia)/. 2010. "Buto: Antica città egiziana del delta (od. Tell el-Farein), nata dalla fusione di due centri urbani affiancati, Pe e Dep. Dell'importanza storica e religiosa della località – che fu il centro di uno dei due grandi Stati (del Nord e del Sud) in cui si divideva l'Egitto predinastico – manca ancora un adeguato riscontro archeologico". https://en.wikipedia.org/wiki/Buto. "Buto … was a city that the Ancient Egyptians called Per-Wadjet. It was located 95 km east of Alexandria in the Nile Delta of Egypt. What in classical times the Greeks called Buto, stood about midway between the Taly (Bolbitine) and Thermuthiac (Sebennytic) branches of the Nile, a few kilometers north of the east-west Butic River and on the southern shore of the Butic Lake (Greek: Βουτικὴ λίμνη, Boutikē limnē). Today, it is called Tell El Fara'in ("Hill of the Pharaohs"), near the villages of Ibtu (or Abtu), Kom Butu, and the city of Desouk (Arabic: دسوق). … Buto was a sacred site in dedication to the goddess Wadjet. It was an important cultural site during prehistoric Egypt (before 3100 BCE). The Buto-Maadi culture was the most important Lower Egyptian prehistoric culture, dating from 4000–3500, and contemporary with Naqada I and II phases in Upper Egypt. … Archaeological evidence shows that the Upper Egyptian Naqada culture replaced the Buto-Maadi culture (also known as the "Lower Egyptian Cultural Complex"), perhaps after a conquest. More recently, scholars have expressed reservations about this; they pointed out that, in the Delta, there was a considerable transitional phase. The unification of Lower Egypt and Upper Egypt into one entity is now considered to be a more complex process than previously thought. … In the earliest records about the region, it contained two cities, Pe and Dep. Eventually, they merged into one city that the Ancient Egyptians named Per-Wadjet. The goddess Wadjet was the patron deity of Lower Egypt and her oracle was located in her renowned temple in this area. An annual festival was held there that celebrated Wadjet. The area also contained sanctuaries of Horus and Bast, and much later, the city became associated with Isis. … The patron deity of Lower Egypt, Wadjet, was represented as a cobra. The patron deity of Upper Egypt, Nekhbet, was depicted as a white vulture. Their separate cultural statures were such essential features that they never merged when the two cultures unified. … During the Ptolemaic Kingdom, a Greek-speaking dynasty that ruled from 305 to 30 BCE, the Greeks coined the toponym Buto for the city. It served as the capital, or according to Herodian, merely the principal village of the Delta. … Greek historians recorded that Buto was celebrated for its monolithic temple and the oracle of the goddess Wadjet (Buto), and a yearly festival was held there in honour of the goddess. …". Then, was it the Buto shrine? Is it the temple of Wadjet?

Returning to Shepseskaf's mastaba. Wikipedia:

> The narrow ends of the mastaba were deliberately raised unlike the traditional fashion, making the tomb look like a great sarcophagus[3][158] or the hieroglyphic determinative for a shrine.[note 24][142] The mastaba was originally clad with white Turah limestone except for its lower course, which was clad in red granite.[41][157] The entrance to the

https://en.wikipedia.org/wiki/Shepseskaf#cite_note-183 note 24: "That is, the mastaba took the shape of a Buto shrine with a rounded vaulted top between vertical ends.[157]". The reference is Lehner 2008, p. 139. "Menkaure's successor, Shepseskaf, chose to be buried in South Saqqara, under a huge mastaba, 99.6 m (327 ft) long by 74.4 m (244 ft) broad, originally encased with fine limestone, except for a bottom course of red granite. With an outer slope of 70°, *it may have risen in two steps and certainly took the form of a Buto shrine* — a vaulted top between vertical ends. A corridor descends at 23° 30' for 20.95 m (69 ft) to a corridor-chamber followed by three portcullis slots and a passage to an antechamber. A short passage slopes down to the west to the burial chamber. Its ceiling, like Menkaure's, was sculpted into a false vault. Remains were found of a hard dark stone sarcophagus, decorated like Menkaure's (p. 135). From the southeast of the antechamber a narrow corridor leads to six niches, the equivalent of those in the tombs of Menkaure and Khentkawes, and the precursor of the three small magazines that would become standard. The mastaba was surrounded by a double enclosure defined by mudbrick walls. A small mortuary temple on the east had an offering hall and false door, flanked by five magazines. There were no statue niches though part of a statue of Shepseskaf was found in the temple. To the east lay a small inner court and a larger outer one. A long causeway led to a valley temple which has never been excavated" (Lehner, 2008)

**The form of the shrine**

In Magli, 2024, about the form of the tomb, it is told the following: "the tomb resembles a specific, sacred building: a "Buto shrine" (Lehner 1999). The form of these archaic edifices … is known from the corresponding hieroglyph, representing an arched roof building with side poles". Magli gives O20 Gardiner sign

https://en.wikipedia.org/wiki/List_of_Egyptian_hieroglyphs

Magli continues: "Buto was an important cult centre, located in the delta. The Cobra-goddess Wadjet …".

About the hieroglyph, and besides Wikipedia, we can propose Abd-Alghafour et al., 2023, and their work about the "Representation of 'Saw' and 'Pr-nw' Shrines in the Funeral Procession in Some Non-Royal Mastabas During the Fifth and Sixth Dynasties." "This study sheds light on the symbolic depiction of both Sais and Buto shrines, whether they were depicted by the kings or later by the individuals, such as viziers and high officials. It deals with the subject from the second half of the Fifth Dynasty to the Sixth Dynasty, especially during the reigns of Djedkare Isesi and Unas kings. These shrines were depicted in the funeral procession as part of a river trip to the sacred religious shrines before the completion of the burial. The study depends on seven non-royal mastabas in Saqqara and Dahshur as sources for the visits to Sais and Buto shrines. It also attempts to link that visit to the high social and administrative status of the viziers and high officials. Last but not least, the study attempts to indicate the shrines' number, symbolism, building material, and architectural features in addition to their religious and funeral roles" (Abd-Alghafour et al., 2023).

https://jarch.journals.ekb.eg/article_277032_65417a644d0b0b2afb3cd22f2ce2b4dc.pdf

Here a screenshot, where we find mentioned the Buto shrines, and we can also see the associated hieroglyphs (Abd-Alghafour et al., 2023):

> Buto played an essential role in religious and funeral rites at the beginning of the Pre-Dynastic Period as it was the center of the capital and the cemetery of the ancient rulers of the delta.[79] Meanwhile, both shrines, 'pr-nw' and 'pr-nsr' of Lower Egypt had the primitive design of the ancient Egyptian temple with vaulted ceilings symbolizing the two districts of Buto, 'P' and the 'Dp'.[80] The two shrines were depicted on the *Scorpion* King's macehead from the Pre-Dynastic Period.[81]

"When the shrines of Buto were depicted in religious rituals, they actually referred to the temple in general. In contrast, when they were depicted in the funeral rites, they symbolized the cemetery and the tombs of the ancient Buto rulers, where the 'mww' dancers who represent ancient Buto rulers go out to receive the funeral procession. Buto also had significance in the afterlife as the deceased wished to be buried in the shade of the sacred grove in Buto to be satisfied and blessed. Besides, it is the area where the goddess Isis gave birth to her child Horus" (Abd-Alghafour et al., 2023, and references therein).

"The royal religious visits to the symbolic shrines of the holy cities continued from the Third Dynasty up to the Late Period. The best examples of these visits can be seen in the 'Hb-sd' celebrations during which the king is represented symbolically dead and then resurrects so both his power and legitimacy are renewed post his symbolic visits to the shrines of the deities of Upper and Lower Egypt. The king's pyramid complexes of Djoser, Sneferu, Sahure, Niuserre, Pepi II, Senusret III and Amenemhat Sobekhotep I are great attestations of such depictions" (Abd-Alghafour et al., 2023, and references therein).

**Akhet Ra**

Let us repeat what we have read in Sellers, 1992. "About 2473 BC we find sun temples built some miles to the South of Giza. Six rulers in the Fifth Dynasty built these great temples to the sun and named them such names as Pleasure of Re, Horizon of Re and Field of Re".

Horizon of Ra is Akhet Ra.

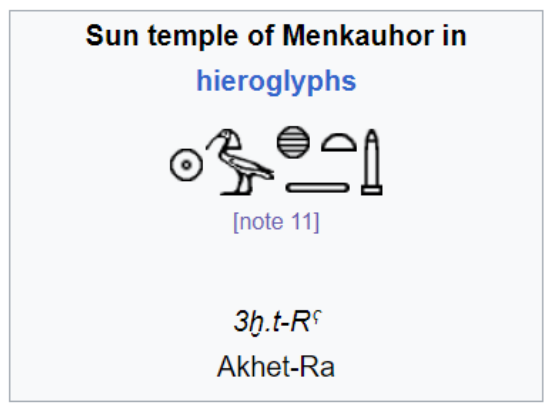

THE CRESTED IBIS! AKHET! https://en.wikipedia.org/wiki/Menkauhor_Kaiu . Note 11: "The last hieroglyph shown here is an approximation of the correct one which shows a squat obelisk on a flat base called a ben-ben." Then, from right to left, the temple of Akhet Ra.

**Mastaba of Qar**

From Chris Naunton web site (2024): we find a slide from Naunton's talk "showing one of the titles held by the official Qar and containing the name of the Great Pyramid, the 'Horizon of Khufu' which is written with a pyramid determinative sign." Here we show a part of the slide:

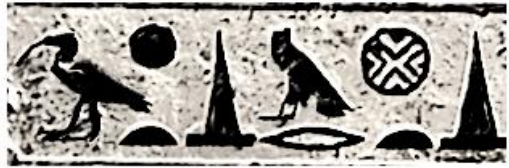

It is written: "Overseer of the town of the pyramid Akhet", then it follows the Khufu's cartouche. Note the crested ibis. Note the determinative "pyramid". Note the semicircle for the letter "t", known as the "bread bun".

In Simpson, 1976, we can find described the mastabas of Qar and Idu, G7101 and 7102. gizamedia.rc.fas.harvard.edu .

> B 2  Imy-r niwt Ȝht-Ḥwfw sš ʿ nswt ḫft ḥr imȝḫ(w) Ḳȝr,
> "overseer of the pyramid city Akhet-Khufu, king's letter
> scribe in the presence, the well provided Qar."

"Mr. Ashraf Mohieddin, General Manager of Antiquities of the Pyramid, said that the restoration work in the tomb of Qar included mechanical cleaning of the entire tomb, strengthening the roof at the entrance, and filling the gaps in the upper part of the main hall walls. It is a rock-cut tomb dating to the Sixth Dynasty, the reign of King Pepi I. Qar was the overseer of the two pyramid towns of Khufu and Menkaure, the inspector of the wab (pure) priests of the pyramid of Khafre, and the tenant of the pyramid of Pepi I." (https://egymonuments.gov.eg/news/opening-of-the-tombs-of-edo-and-qar/)

"As for the tomb of Idu, Mohieddin indicated that … This rock-cut tomb dates to the reign of the Sixth Dynasty King Pepi I as well. Idu was the son of Qar. His titles were the tenant of the Pyramid of Pepi I, the overseer of the scribes, and the inspector of the wab (pure) priests of the pyramid of Khufu and Khafre." (https://egymonuments.gov.eg/news/opening-of-the-tombs-of-edo-and-qar/)

**Akhet Khufu**

In Magli, 2024, we find: "Khufu will indeed build his pyramid on the Giza plateau, in plain view from Heliopolis (the main theological centre of the cult of Ra), and will make an explicit reference to the sun cult with the spectacular hierophany occurring at Giza at the summer solstice re-creating the "solarized" version of the double mountain sign, Akhet … (Lehner 1985, Shaltout et al. 2007, Magli 2008)".

The Fig.13 in Shaltout et al., 2007, shows the "hierophany", corresponding to the "solarized" horizon, that is the picture of the sun setting between the two pyramids.

The hieroglyph is Gardiner N27. https://en.wikipedia.org/wiki/Gardiner%27s_sign_list

| N27 | 𓈌 | U+1320C | sun rising over mountain | "horizon" |

Here in the Appendix, I repeat what I wrote in https://arxiv.org/pdf/2411.08061, adding a further important observation about the "bread bun". The reader can find the discussion on Page 11 of that work. In my article, 2411.08061, the reader can understand the fundamental importance of the Wadi

al-Jarf papyri, discovered by Pierre Tallet, 2017. **The papyri are of Khufu's time**: SEE PLEASE THE PAPYRUS at the-past.com and note the AKHET below the Khufu's cartouche. Akhet is written with the crested ibis.

Here Fig.2, in https://arxiv.org/pdf/2411.08061

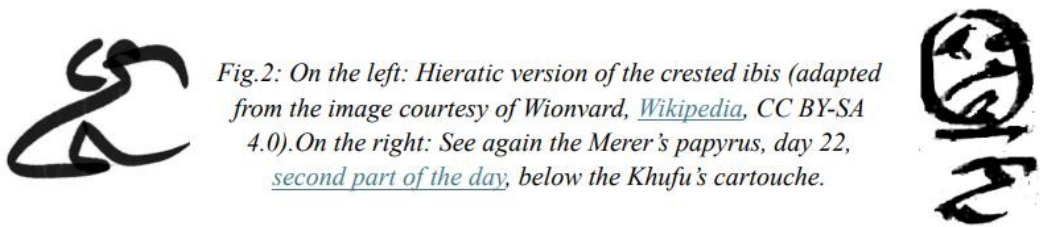

Here the link to the picture, in the article "Records of the pyramid builders: discovering eye-witness accounts of a legendary construction project", The Past., 2023. The picture is a Pierre Tallet courtesy.

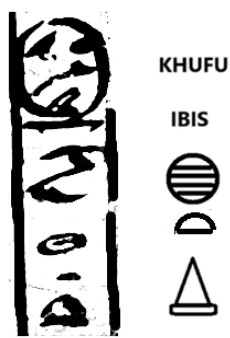

Transcription according to Tallet, 2017.

The combination Aa1 + t (or Aa1 + X1) : https://en.wikipedia.org/wiki/Branch_(hieroglyph)

As the value (kh)t, it is often complemented in a hieroglyphic block with *kh*–("sieve"),

and *"t"*–(bread bun).

The semicircle is the "bread bun". It is the alphabetic-t. It is also used for the feminine determinative.

This is the AKHET that KHUFU knew, AKHET with the CRESTED IBIS; not the horizon with a disk between two hills. Magli must show a text of Khufu's time where AKHET is written with a disk and two hills, not a picture of two pyramids and the sun setting between them. The burden of finding the use of Gardiner list N27 symbol at the time of Khufu's is with Magli.

Let us note that, in 2014, https://arxiv.org/pdf/1401.0508, Magli wrote: "We actually do not know the way in which the name of the Khufu complex was written during Khufu's reign (the unique possible attestation comes from a re-used fragmentary relief where the pyramid's determinative is not present); the "double mountains plus sun" sign first appears during the 5th dynasty". Magli did not provide any reference about the fact that Gardiner N27 appeared during the fifth dynasty.

*Once more, today we have the Wadi el-Jarf papyri. And we know very well how the name of the complex was written during the Khufu's reign. It was written with the crested ibis.*

Let us consider also the discussion by Bauval, 2013. At the archived document, it is possible to see a picture of the supposed 'hierophany', the sun between two pyramids, first proposed by Mark Lehner in 1985, and reproposed by Magli. Bauval is discussing Magli and his arXiv, 2007, where we find Gardiner N27 for Akhet, not the Akhet with the crested ibis, which is the symbol actually used for Akhet Khufu at the time when the pyramid was built (Tallet, 2017). In arXiv 2014, Magli added also the ibis-sign (it is mentioned the mastaba of Qar) but the so-called 'Khufu's project' of the two pyramids remains the same. That is, Magli's proposal is that Khufu planned two pyramids, those of Khufu and Khafre. Khafre, son of Khufu, built the second-largest pyramid at Giza. The name of the pyramid was Khafre-Wer which means "Khafre is Great" (Hawass, 2024). In 2024, Magli maintains his supposed project, without mentioning the true written name of the Pyramid, that we find in the Wadi El-Jarf papyri.

**Earliest Known Egyptian Writing**

The earliest known examples of Egyptian writing are from Abydos, on bone and ivory tags, pottery vessels, and clay seal impressions. "The tags, each measuring 2 by 1 1/2 centimeters and containing between one and four glyphs, were discovered in the late 20th century in Tomb U-j of Umm el Qu'ab, the necropolis of the Predynastic and Early Dynastic kings by excavators from the German Archaeological Institute in Cairo led by Günter Dreyer. Tomb U-j may hold the remains of predynastic ruler Scorpion I. The discoveries in Tomb U-j were first published by Dreyer, Ulrich Hartung, and Frauke Pupenmeier in Umm el-Qaab. Volume 1: Das prädynastische Königsgrab U-j und seine frühen Schriftzeugnisse (1998) (historyofinformation.com).

In the Magli's paper of 2014, the author tells that

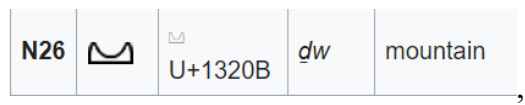,

 a sign he calls djew, "was associated with afterlife and was already extremely ancient, as it appears already in the seals found in the pre-dynastic tombs at Abydos (Dreyer, 1998)". This sentence written by Magli is lacking further detailed information.

As previously told, Dreyer's book, 1998, regards the site of Umm el-Qa'ab, the royal tomb U-j. "Dated to Naqada IIIA2 (3,300 BCE), tomb U-j is the largest tomb found at Cemetery U and contains 12 separate chambers. … Although the tomb had been subject to plundering, about 2000 ceramic vessels were recovered with nearly one third having been imported from Palestine" (Wikipedia). "Ivory tags found at Cemetery U-j: In addition to ceramic vessels, tomb U-j also contained bowls carved out of obsidian and chests made from imported cedar. ... Small ivory tags with hieroglyphics inscribed on them were also recovered from the tomb. These artifacts provide the earliest evidence of writing in Egypt" (Wikipedia, mentioning Stevenson, 2016).

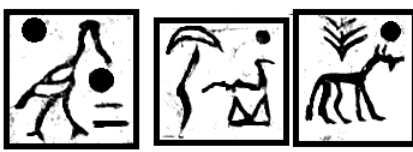  Wikipedia, Ivory tags found at Cemetery U-j (Mitchell, 1999).

Is the sign on a tag (middle picture given above) the dejw? Is it associated with afterlife, besides the fact that it was found in a tomb? Dreyer associates it with the mountains.

To show the tag or seal or seal impression from Umm el-Qa'ab, tomb U-j, which is containing the djew associated with afterlife, is with Magli.

**Elephant on mountains**

According to Magli, there is not only the Khufu's project of two pyramids at Giza; there is also that of the two pyramids of Dashur. In this case, Belmonte is involved too, in the "Global Project of Sneferu", 2015. We find told that, "when viewed from Saqqara, the Sneferu pyramids form an artificial, symbolic horizon of two paired mountains". Symbolic? Being symbolic to Belmonte and Magli, they arrive to the sign:

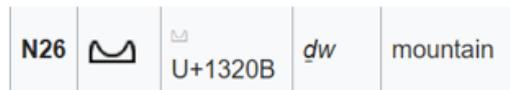

Belmonte and Magli add that "this sign, which is extremely ancient, as it appears already in the seals found in the pre-dynastic tomb U-j at Umm el Qab, the Abydos necropolis, where a 'King Elephant mountain' is mentioned.21 *Its connection with afterlife and with Abydos in particular is clear*. [No reference given. Why is it clears?] To the southern horizon of the necropolis a huge wadi opens into the Umm el-Qa'ab bay, functioning as a sort of 'mouth of the afterworld' [No reference]. The sign appears explicitly in the name of Abydos, … and in curious ritual objects (today called Osiris' reliquary) associated with the Abydos sacred centre [No reference]. Actually, the idea of a 'sign of two mountains' will later be present in several royal funerary landscapes of ancient Egypt, 22 …" (in Belmonte and Magli, the only two references are 21, Dreyer, 1998, and 22, a Magli self-citation).

In Dreyer there is a tag with an elephant (tag, not seal) with protruding tusks, standing on *3 mountain peaks* to the right, next to it is a tree with a bushy, branched crown. Three mountains, not two. We can find it also in Wegner, 2007, in an article entitled "From elephant-mountain to Anubis-mountain? A theory on the origins and development of the name Abdju", available Academia.edu.

"After more than a century of archaeological activity at Abydos, Egyptology has assembled a detailed picture of the evolution of that site. … . Despite the increasing sophistication of our archaeological picture of Abydos, it is perhaps surprising that the origins and meaning of the very name of Abydos during the pharaonic periods—AbDw—remain little understood. A majority of authors remain reticent on the issue of the origin and meaning of this place name. ... A second issue which inevitably presents scholars with pause in discussing the original meaning of the name AbDw is the fact that we possess no certain attestation of the name prior to its appearance in the Old Kingdom. When Abydos first appears in Old Kingdom funerary prayers and in the Pyramid Texts it is already written with variants of the familiar orthography seen in subsequent periods. Was the historical toponym AbDw one that derived ultimately from the Predynastic place-name of the site?  … The vacuum on the early toponyms associated with Abydos has been broken in recent years, stemming largely from the publication of the material from the Nagada IIIA period tomb U-j at Umm el-Gaab. Within the corpus of inscribed tags in that tomb, occur a number bearing an elephant symbol atop a triple-peaked mountain hieroglyph. Günter Dreyer interpreted the text on these tags as an archaic writing of the toponym Ab-Dw, with meaning "Elephant-Mountain." He argued that the historical place-name had its origins in the late Predynastic period through virtue of a king Elephant, a hypothetical predecessor of king Scorpion, the presumed occupant of tomb U-j (Dreyer 1998: 140–141, 173–180). The historical toponym Abydos thus meant in origin something akin to: "Mountain of (King) Elephant."

(Wegner, 2007). Wegner continues with a detailed discussion of further literature and illustrates the local topography of Abydos as representing an elephant. He also proposes, in his Fig.5, the orthographic variations of AbDw. In his Figs. 7 and 8, the reader can find: in Fig.7, "tags from Tomb U-j with use of elephant symbols: Elephant atop triple peaked mountain sign (nos. 1–7) and recumbent "elephant" with Upper Egyptian shrine (nos. 8–13), after Dreyer 1998", and in Fig.8, "tags from Tomb U-j showing use of double and triple peaked mountain symbols (nos 1–4), and tags possibly designating Abydene locales (5–8), after Dreyer 1998" (Wegner, 2007).

Let us suggest also Francesco Raffaele's web site.

As previously reported, Belmonte and Magli, 2015, tell: "To the southern horizon of the necropolis a huge wadi opens into the Umm el-Qab bay, functioning as a sort of 'mouth of the afterworld' [No reference]. The sign appears explicitly in the name of Abydos, …".

From Wegner, 2007, "one of the defining characteristics of Abydos is the way the site grew up to occupy the southernmost section of that embayment. The greater bay that contains Abydos covers a total area of some 3 × 5 km. However, the town and cemetery fields of Abydos through all periods remained largely confined to the southern half of the bay (Fig.1 [in Wegner, 2007]). The early site extends from the position of the ancient settlement core, the Kom es-Sultan, westwards to Umm el-Gaab which occupies a low rise circumscribed by the low-desert wadi that served as natural route of approach to the necropolis. The position of Umm el-Gaab might originally have been related to the high desert wadi behind it as a symbolic western point of entry to the netherworld. Gravitation towards the southern end of the bay undoubtedly was solidified during the pharaonic period by the role of Umm el-Gaab as the symbolic tomb of Osiris with the low-desert wadi forming a ceremonial axis between the Osiris-Khentiamentiu temple precinct and that sacred site. Once this combination of ritual elements was formally established, tendency was for activity over long periods to nucleate tightly around this focal point of the site" (Wegner, 2007, and references therein).

**Two or three hills**

Courtesy https://en.wikipedia.org/wiki/List_of_Egyptian_hieroglyphs

| | N25 U+13209 | three hills | Foreign land (ḫꜣst) Determinative for foreign locations | | N28 U+1320D | rays of sun over hill | Appear (ḫꜥj) |
|---|---|---|---|---|---|---|---|
| | N25A U+1320A | three hills (low) | Foreign land (ḫꜣst) Determinative for foreign locations | | N29 U+1320E | slope of hill | Hill (qꜣꜣ) |
| | N26 U+1320B | two hills | Mountain (ḏw) | | N30 U+1320F | mound of earth | Mound (jꜣt) |
| | N27 U+1320C | sun over mountain | Horizon (ꜣḫt) | | N31 U+13210 | road with shrubs | Path, way (wꜣt) Course, road (mtn) Distant (hr) |

**Returning to Wadi al-Jarf**

Belmonte and Magli mention the Wadi al-Jarf papyri in 2015, in "the global project of Sneferu at Dahshur". In this 2015 paper, Belmonte and Magli are reproposing the idea that the two largest pyramids at Giza were planned in a single project by Khufu. Referring to "The harbour of Khufu on the Red Sea coast at Wadi al-Jarf, Egypt", Near Eastern Archaeology, Belmonte and Magli tell that Tallet and Marouard "have found a papyrus, contemporaneous to Cheops's reign, with a text where Axt xfw is mentioned with a single determinative of a pyramid (as in later sources). This, according to José Lull (private communication), would be a point against the duality of the complex". Please consider the article by Lehner about Merer and the Sphynx. In a book by Belmonte and Lull, 2023, "Astronomy of Ancient Egypt", we can find in the Introduction, mentioned "Taller" and Lehner, 2021. Belmonte and Lull, 2023, are just telling that surprises are always possible as "demonstrated by the recent discovery of papyrus fragments of Khufu's pyramid accountancy in a barren spot of the Red Sea coast (Taller & Lehner, 2021)". Nothing more. The "barren spot" is Wadi al-Jarf that hosted the artificial harbor of Khufu, where phyles linked Egypt to Sinai.

There are several papyri found at Wadi al-Jarf, not only one papyrus.

***Let us stress that Akhet Khufu is written with the crested ibis and not with the double mountain sing. The name of the Khufu's Pyramid is clearly shown in the Wadi al-Jarf papyrus, known as the Diary of Merer (Tallet, 2017).***

**Assmann and Lehner**

Magli, 2024: "Indeed the interplay between symbolic hieroglyphs and architecture is typical of the way the Egyptians conceived writing, as magisterially put in evidence by Assmann (2007). According to Assmann the very idea of creation was tied up with writing in the Egyptian's mind; in a sense the pyramid was a "sign" within a "written" sacred landscape. With the IV dynasty the pyramids become "gigantic hieroglyphs", as Lehner (1999) puts it; first the double-mountain sign with Snefru [Sneferu], and then an icon of glory – Akhet, the Sun between the two mountains – with Khufu (Magli 2016, [preprint 2014](#))." (Magli, 2024).

About Magli referring to Lehner and Assmann, let us read Lehner and what he wrote. "The Pyramid as Icon". "The pyramid was above all an icon, a towering symbol. It has been said [who and where? Lehner did not provide references] that the Egyptians did not distinguish sharply between hieroglyphic writing, two-dimensional art and relief carving, sculpture and monumental architecture. In a sense, *the pyramids are gigantic hieroglyphs*. But why a pyramid? And how should we read the pyramid glyph? Pyramid and pyramidion. The word for pyramid in ancient Egyptian is met. There seems to be no cosmic significance in the term itself. I.E.S. Edwards, the great pyramid authority, attempted to find a derivation from m, 'instrument' or 'place', plus ar, 'ascension', as 'place of ascension'. Although he himself doubted this derivation, the pyramid was indeed a place or instrument of ascension for the king after death. … The capstone or pyramidion is the complete pyramid in miniature, bringing the structure to a point at the same angle and with the same proportions as the main body. Stadelmann found the earliest pyramidion at Sneferu's North Pyramid at Dahshur, made of the same limestone as the casing and uninscribed" (Lehner, 1999)

Then, let us read Creighton, 2014.

"Professor Jan Assmann writes: *In Egyptian the pyramid of Cheops (whose Egyptian name was Khufu) is called Akhet Khufu. Akhet is the threshold region between the sky, the earth, and the underworld; in particular, Akhet is the place where the sun rises. The etymological root of the word has the meaning of 'blaze, be radiant'; likewise, the hieroglyph for akhet has nothing in common with the pyramid, but is a pictogram of the sun rising or setting between two mountains. The pyramid does not represent such an akhet, but symbolizes it in an aniconic way. The term of comparison between akhet and pyramid is the idea of 'ascent to heaven'. As the sun god ascends from the Underworld to the akhet and appears in the sky, so the king interred in the pyramid ascends to heavens by way of his akhet, his threshold of light*" (Creighton reporting Assmann).

Akhet, at the time of Khufu's reign was written with the CRESTED IBIS, not with the sign of the sun between two mountains. In Lehner, the pyramid is an icon, in Assmann, it becomes an "aniconic way".

Creighton, 2014, notes that what told by Assmann is all very symbolic. Egyptologists interpret the pyramid as the place where the king became an akh. "However, this interpretation of Akhet Khufu put forward by Assmann is all well and very good except for the not-insignificant problem that the akhet pictogram for "horizon" (believed to depict the sun rising between two mountains) did not actually exist when Khufu was building his Great Pyramid, as is implied in the Assmann's quote above. Indeed, this pictogram only came into being around the end of the Fifth Dynasty, long after Khufu and the completion of the early, giant pyramids" (Creighton, 2014).

"Notwithstanding this inconvenient fact, the early Old Kingdom of Egypt used, according to Egyptologists, a different version of the word akhet (interpreted by many Egyptologists also as meaning 'horizon'). The pictogram for this supposedly earlier version of the word akhet (horizon) is enterely different from the sun disc between two mounds and invokes instead the use of the ibis, which has various translations, inter-alia, 'intellicenge', 'illuminatio', 'shining', 'beneficial'. And 'useful'" (Creighton, 2014).

"supposedly earlier version"? We are sure that akhet is written with the ibis at the time of Khufu's reign!

Creighton, 2014, mentions Lehner. "Lehner was probably one of the first academics to recognize the translation problem that these two quite distinct versions of the word akhet present. In Lehner's view the older term akhet, with the crested ibis, should not be translated as 'horizon' at all but instead is to be associated with the 'Spirit of Khufu'. Lehner writes, 'joining the stars, the king becomes an akh, [and] Akh is often translated as 'spirit' or 'spirit state'. It derives from the term for 'radiant light', written with the crested ibis … Akh is also the word for 'effective, 'profitable', 'useful'" (Creighton, 2014.

Then, let us repeat **Lehner, 2001**. "Akhet is usually translated as 'horizon', where land and the skies touch, but it meant much more in the Egyptian world concept. Written with the same root as the word akh, the Akhet was where the dead were transformed into effective inhabitants of the world beyond death. As part of the sky, it was also the place into which the sun, and therefore the king, was reborn from within the Dua. It is not hard to imagine the early Egyptians being inspired by the pre-dawn glow in the eastern horizon, and by the sunset flaming in the west, to see the area just below the horizon as the place of glorification. Khufu's pyramid was Akhet Khufu. ***Here, and in the Pyramid Texts, Akhet is written with the crested ibis and elliptical land-sign, not with the hieroglyph of the*** *sun disk between two mountains that was used later to write 'horizon'*. As the place where the deceased becomes an akh, a suggested translation is 'Spirit' or 'Light Land'."

**Note please that Akhet Khufu was written in the Wadi al-Jarf and in the Qar inscription without the sign of the 'elliptical land sign' N18. Lehner, 2001, was wrong.**

See please the Supplementary material in "The Egyptian Hieroglyph of the Crested Ibis, from the Cheops' pyramid (Akhet Khufu) to the Akhenaten's Glory of Aten", available https://doi.org/10.5281/zenodo.14587114. There I proposed Akhet Khufu as the "Enduring Forever Glory of Khufu".

**The so-called "horizon"**

Kuentz, 1920, wrote «Autour d'une conception égyptienne méconnue: l'Akhit ou soi-disant horizon », About a little-known Egyptian concept: the Akhet or so-called horizon.

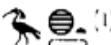

The origin is in Champollion: "Montagne Solaire". Kuentz discussed in depth the literature. Note that Kuentz shows also the 'elliptical land' N18.

Kuentz considers the inscriptions which contain Akhet. So please look at Kuentz, for the details. He observes that "pyramid" is masculine,

Let us remember that the 'bread bun' is letter "t" and the feminine determinative. We find the bread bun in the name of a man, Akhethetep.

Kuentz used the sign N18 of the 'horizon'. Today, we have the Merer's diary on the Wadi al-Jarf papyri. Once more, please observe the image in The Past, which is giving a fragment of Papyrus B. In the transcription by Pierre Tallet, 2017, there is the bread bun, not N18.

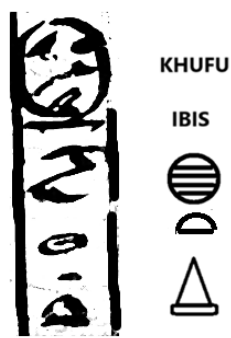

The pyramid Akhet Khufu in Wadi el-Jarf papyri, as given in the picture of the article "Records of the pyramid builders: discovering eye-witness accounts of a legendary construction project", The Past., 2023. The transcription is given according to Tallet, 2017.

In Kuentz, 1920, we can find that the ibis sign in Akhet was used by Pepi II in a letter to his vizier Herkhuf (Harkhuf):

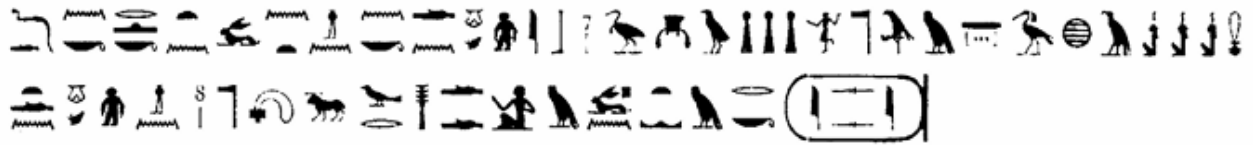

Tu as dit en cette tienne missive que tu amenais de la Terre des Horizontaux un Deng dansant le dieu, semblable au Deng que le scelleur de dieu Ba-our-zeded amena de Pount au temps d'Asesi.

Herkhuf visited the land of the Horizon dwellers (Goedicke, 1981, Bauval, & Brophy, 2011).

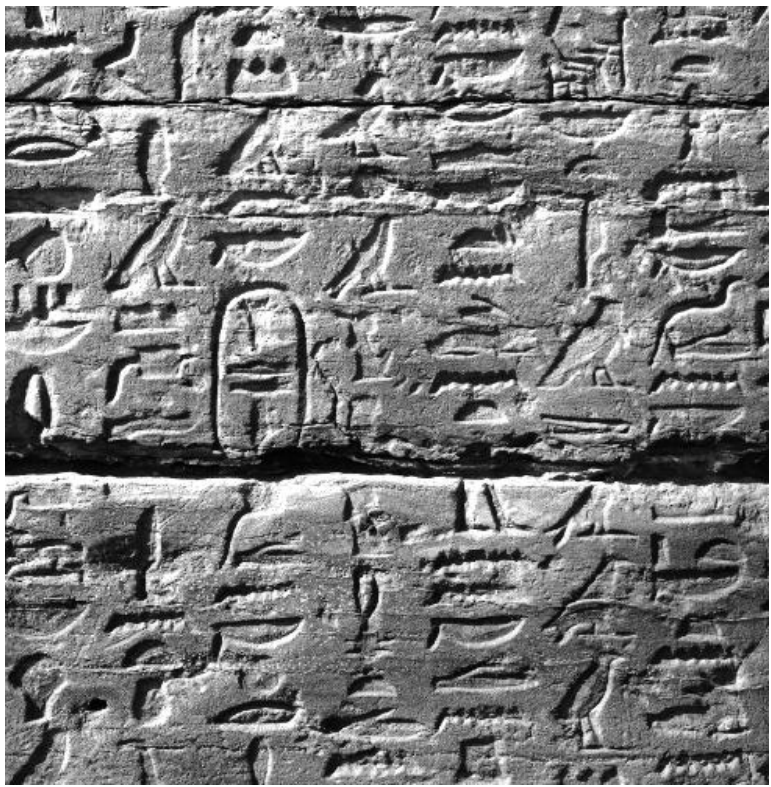

Thanks to the Museo Egizio di Torino, we can see the original text. The images are provided Under CC0 license, "No Rights Reserved". The caption of Photo 10/27 is: "Right wall of the façade of the tomb of Herkhuf (QH34n). The text is a letter from the young Pharaoh Pepi II, in which Herkhuf is honored for his expedition, from where he brought back a pygmy. Schiaparelli excavations." Year, 1914, Palte C01547.

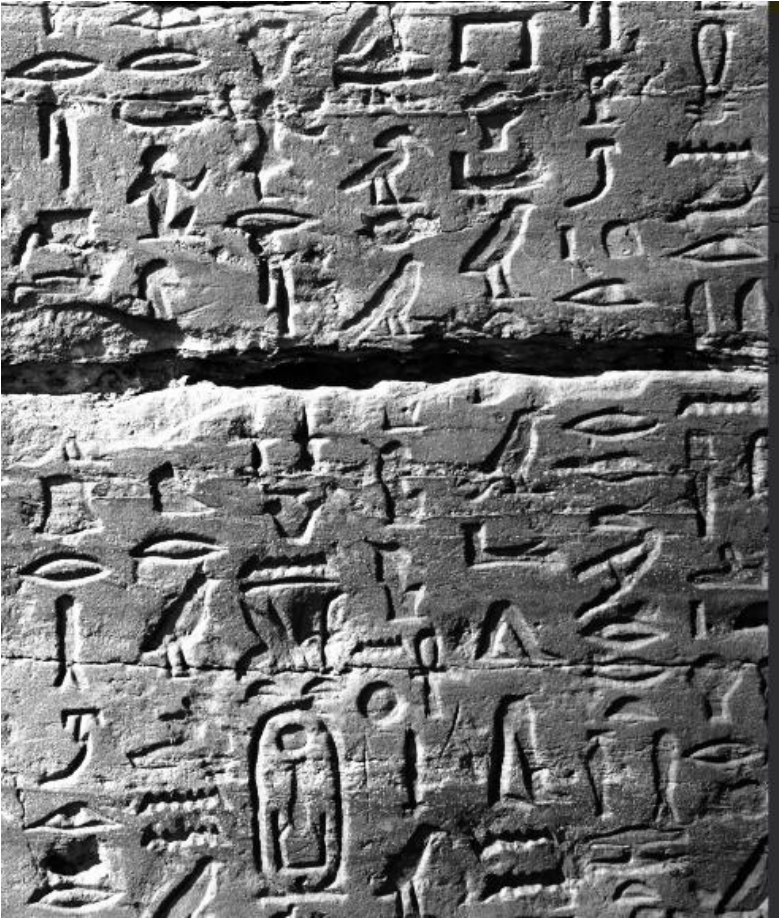

Thanks to the [Museo Egizio di Torino](), we can see the original text of the letter. The images are provided Under CC0 license. Photo 11/27. Schiaparelli excavations." Plate C01548.

In Gautschy et al., the chronology of Pepi II, is the following:

| 1 Pepy II | 2201 BCE | 2216 BCE | 2278 BCE | 2279/2229 BCE | --- | 2334 BCE |

Pepi II is of the sixth dynasty. Then, when N27 has been introduced for "horizon"?

Creighton, 2014, notes that "the akhet pictogram for "horizon" (believed to depict the sun rising between two mountains) did not actually exist when Khufu was building his Great Pyramid, as is implied in the Assmann's quote above. Indeed, this pictogram only came into being around the end of the Fifth Dynasty, long after Khufu and the completion of the early, giant pyramids" (Creighton, 2014). According to the Letter of Pepi II, we can find that the crested ibis for 'horizon' was used even three centuries after Kufhu.

**Appendix - The Akhet environment**

Hong-Quan Zhang, 2023, has investigated the Akhet environment, as we can find depicted by the Pyramid Texts, comparing it with the local hydrology during the Green Sahara time. "The Pyramid

Texts contain vivid, specific, and consistent details of the Akhet and its surrounding lakes, canals, and farmlands" (Zhang, 2023). Using a climate change approach and hydrological profiles of the Green Sahara time, Zhang's "insight provides a key to decipher many obscure words in the current English translations. It also helps pinpoint the scattered environmental descriptions in the Pyramid Texts into a coherent portrait". The article "examines the relevant original hieroglyphs in the Pyramid Texts against their corresponding transliterations and English translations".

Zhang writes:

> The hieroglyph Akhet (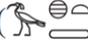 or 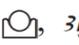, ꜣḫt, meaning "place of becoming effective") is generally translated as "horizon" which blanks its original true identification.

Akhet Khufu is written, as clearly shown by the Wadi el-Jarf papyri, with the crested ibis. The other sign, Gardiner list N27, resembles the sun between two hills.

Robert Bauval, 2013, available Academia.edu, explained that, looking at sign N27 Mark Lehner in 1985 wrote: *[a] dramatic effect is created at sunset during the summer solstice as viewed, again, from the eastern niche of the Sphinx Temple. At this time, and from this vantage, the sun sets almost exactly midway between the Khufu and Khafre pyramids, thus construing the image of the Akhet, 'Horizon', hieroglyph on a scale of acres*". Mentioning archaeoastronomer G. Magli, Bauval tells that it has been concluded by Magli that Khufu and Khafre pyramids "were deliberately positioned relative to each other to create the hieroglyphic sign N27 'Akhet' i.e. "sun disk between two mountains", and this is why "Khufu called this project 'Akhet Khufu'" (Bauval, 2013). Bauval continued explaining that, "however, a *fatal flaw* [exists] with this idea: the hieroglyphic sign of the 'sun disk between two mountains' did not exist when Khufu built his pyramid! And even if it did, it was not used in the writing of the name 'Akhet Khufu'. In 1997 Lehner did acknowledge this fact: *Khufu's pyramid was Akhet Khufu. Here, and in the Pyramid Texts, Akhet is written with the crested-ibis and elliptical land-sign, not with the hieroglyph of the sun disk between two mountains that was used later to write 'horizon'.*" (Bauval, 2013).

Please consider the Lehner's words in the Lehner's book: "Khufu's pyramid was Akhet Khufu. Here, and in the Pyramid Texts, **Akhet is written with the crested ibis and *elliptical land-sign*, not with** the hieroglyph of the sun disk between two mountains that was used later to write 'horizon'. As the place where the deceased becomes an akh, a suggested translation is 'Spirit' or 'Light Land'." (Lehner, The Complete Pyramids, 1997, reprinted 2001).

***Please consider that, in the Wadi al-Jarf papyri, the Akhet is written with the "bread bun" (Tallet, 2017) and not with the N18. Lehner, 1997, we wrong. We discussed this subject in detail in "The Egyptian Hieroglyph of the Crested Ibis, from the Cheops' pyramid (Akhet Khufu) to the Akhenaten's Glory of Aten", https://doi.org/10.5281/zenodo.14587114. There I proposed Akhet Khufu as the "Enduring Forever Glory of Khufu".***

Robert Bauval stresses that "sign N27 is not found *in the Pyramid Texts*, where the word 'Akhet' is

written with the crested-ibis sign G25 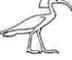, and the elliptical land-sign N18 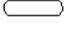. The same applies for the word 'Akhet' in the name 'Akhet Khufu', where 'Akhet' is also written with the crested-ibis sign G25 and elliptical land-sign N18. ... ". Bauval, published 2013, wrote surely before the pictures of the Wadi al-Jarf papyri of the Merer' diary had been available. Let us stress that in Akhet Khufu N18 is absent.

"Amazingly, this mistake in interpretation of the name of the Great Pyramid is still found in recently published books. To be fair, Lehner [in his The Complete Pyramids, 1997, p. 29] did recognize that 'Akhet' - although erroneously translated as 'horizon' in many books - is written with the crested-ibis that also denotes the word Akh, meaning a 'spirit' who lives in the Duat (afterworld), the latter often written with a star in a circle". Bauval, 2013, continues adding that Lehner offered another meaning for 'Akhet Khufu', that is, *the place where the deceased (king Khufu) becomes an Akh, a suggested translation is "Spirit" or "Light Land"*. According to James P. Allen, who translated the Pyramid Texts, "The Axt (Akhet) is the place in which the king, like the sun and other celestial beings, undergoes the final transformation from the inertness of death and night to the form that allows him to live effectively - that is as an akh - in his new world. It is for this reason that the king and his celestial companions are said to "rise from the Axt (Akhet)," and not because the Axt (Akhet) is a place on the horizon or - as some have suggested - because it is a place of light.' Akhet, therefore, must mean the 'Place of Becoming Akh'..." (Bauval, 2013). This is the interpretation provided by J.P. Allen, as reported by Robert Bauval.

Please consider the discussion provided by Bauval. At the archived document, it is possible to see a picture of the supposed 'hierophany', the sun between two pyramids, first proposed by Mark Lehner in 1985, and reproposed by Magli. Bauval is discussing Magli and his arXiv, 2007, where we find Gardiner N27. In arXiv 2014, Magli added also the ibis-sign (it is mentioned the mastaba of Qar) but the so-called 'Khufu's project' of the two pyramids remains the same. That is, Magli's proposal is that Khufu planned two pyramids, that is that of Khufu and Khafre. Khafre, son of Khufu, built the second-largest pyramid at Giza. The name of the pyramid was Khafre-Wer which means "Khafre is Great" (Hawass, 2024).

For what is regarding the "dramatic effect", which "is created at sunset during the summer solstice", let us stress that the colossal statue of the Sphynx is facing east, and therefore not the sunset for sure. Please consider reading carefully the article by Lehner, 2020, entitled "Merer and the Sphinx".

Here is the conclusion of 2020 Lehner's work. "In our friendly dispute with our mentor, Rainer Stadelmann, Zahi [Hawass] once asked, "If Khufu made the Great Sphinx, what was its purpose?" This is a good question. Khufu's causeway extended some 800 m from his upper temple and swung far to the northeast to join his Valley Temple. The only possible reason for Khufu to carve the colossal Sphinx statue here would have been along the lines of the Colossus of Rhodes, or other harbor statues, the Statue of Liberty for an example in our time, in order to monumentalize the gateway to the Giza Necropolis, to the extent that city of the dead was under development flanking Khufu's Pyramid, as the pyramid itself rose (Jánosi 2005, 2006). These "elite" cemeteries were probably not accessible via the Khufu Valley Temple and causeway. Those courtiers who had been assigned the high-status mastabas must have ascended to the cemetery from the proposed landing place of Merer, at the end of Giza's central canal basin. But where else in Egypt do we find such a colossal image separated from a royal tomb or temple complex, save, perhaps, at Abu Simbel? True, like the Abu Simbel colossi, the Sphinx was part of a temple complex, and created as part of a quarry, construction, and image-shaping process. What is as certain as anything can be in archaeology, by virtue of the structural stratigraphy and by the Baugeschichte [building history] of the Sphinx, Khafre Valley Temple, and Sphinx Temple: It was Khafre's builders who carried out the greater part of that process and completed the Sphinx" (Lehner, 2020, and references therein).

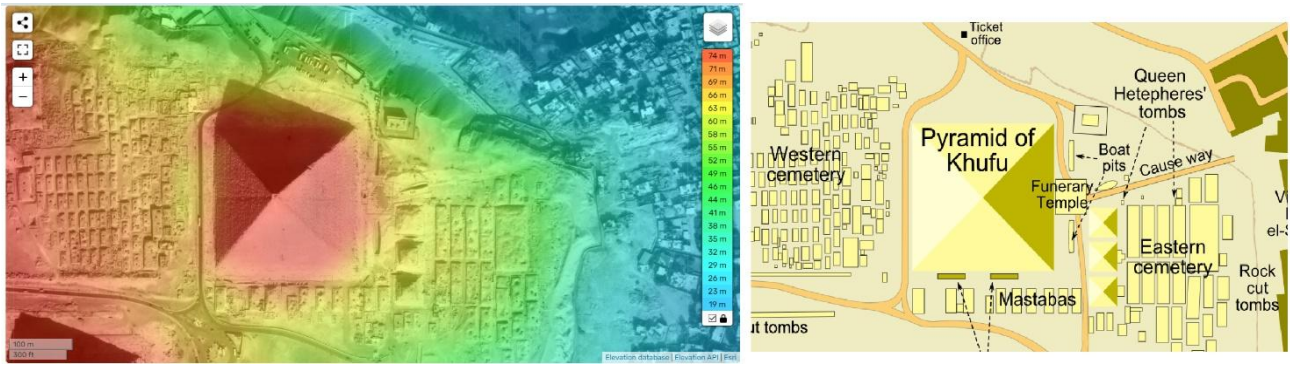

*Akhet Khufu in TessaDEM. On the right a detail oft he map courtesy MesserWoland, CC BY-SA 3.0.*